\documentclass[twocolumn,tighten,times]{aastex631}
\usepackage[utf8]{inputenc}
\usepackage{enumitem}
\usepackage{amsmath}
\usepackage{graphicx}
\usepackage{color}
\usepackage{url}
\usepackage{hyperref}
\usepackage{ulem}

\shorttitle{The green valley galaxies}
\shortauthors{Mahoro et al.}

\graphicspath{{./}{Outflowsplot/}}

\begin{document}

\title{The [OIII] profiles of far-infrared active and non-active optically-selected green valley galaxies}
\correspondingauthor{Antoine Mahoro}
\email{antoine@saao.ac.za}

\author[0000-0002-6518-781X]{Antoine Mahoro}
\affiliation{South African Astronomical Observatory (SAAO), P.O. Box 9, Observatory 7935, Cape Town, South Africa}
\affiliation{Department of Astronomy, University of Cape Town (UCT), Private Bag X3, Rondebosch 7701, South Africa}

\author[0000-0001-7673-4850]{Petri V\"ais\"anen}
\affiliation{South African Astronomical Observatory (SAAO), P.O. Box 9, Observatory 7935, Cape Town, South Africa}
\affiliation{Southern African Larger Telescope (SALT), P.O. Box 9, Observatory 7935, Cape Town, South Africa}

\author[0000-0002-9766-6110]{Mirjana Povi\'c}
\affiliation{Astronomy and Astrophysics Research and Development Department, Entoto Observatory and Research Center (EORC), Space Science and Geospatial Institute (SSGI), P.O. Box 33679, Addis Ababa, Ethiopia}
\affiliation{Instituto de Astrof\'isica de Andaluc\'ia (IAA-CSIC), Glorieta de la Astronom\'ia s/n, 18008 Granada, Spain}
\affiliation{Physics Department, Faculty of Science, Mbarara University of Science and Technology (MUST), P.O. Box 1410, Mbarara, Uganda}

\author{Pheneas Nkundabakura}
\affiliation{MSPE Department, School of Education, College of Education, University of Rwanda (UR),  P.O. Box 5039, Kigali, Rwanda}

\author{Kurt van der Heyden}
\affiliation{Department of Astronomy, University of Cape Town (UCT), Private Bag X3, Rondebosch 7701, South Africa}
\affiliation{National Research Foundation (NRF), P.O. Box 2600, Pretoria, 0001, South Africa}

\author[0000-0002-7705-2525]{Sara Cazzoli}
\affiliation{Instituto de Astrof\'isica de Andaluc\'ia (IAA-CSIC), Glorieta de la Astronom\'ia s/n, 18008 Granada, Spain}

\author{Samuel B. Worku}
\affiliation{Astronomy and Astrophysics Research and Development Department, Entoto Observatory and Research Center (EORC), Space Science and Geospatial Institute (SSGI), P.O. Box 33679, Addis Ababa, Ethiopia}

\author[0000-0003-2629-1945]{Isabel M\'arquez}
\affiliation{Instituto de Astrof\'isica de Andaluc\'ia (IAA-CSIC), Glorieta de la Astronom\'ia s/n, 18008 Granada, Spain}

\author[0000-0002-3170-4137]{Josefa Masegosa}
\affiliation{Instituto de Astrof\'isica de Andaluc\'ia (IAA-CSIC), Glorieta de la Astronom\'ia s/n, 18008 Granada, Spain}

\author[0000-0001-5373-6669]{Solohery M. Randriamampandry}
\affiliation{South African Astronomical Observatory (SAAO), P.O. Box 9, Observatory 7935, Cape Town, South Africa}
\affiliation{Southern African Larger Telescope (SALT), P.O. Box 9, Observatory 7935, Cape Town, South Africa}
\affiliation{A\&A, Department of Physics, Faculty of Sciences, University of Antananarivo, B.P. 906, Antananarivo 101, Madagascar}

\author{Moses Mogotsi}
\affiliation{South African Astronomical Observatory (SAAO), P.O. Box 9, Observatory 7935, Cape Town, South Africa}
\affiliation{Southern African Larger Telescope (SALT), P.O. Box 9, Observatory 7935, Cape Town, South Africa}

\begin{abstract}
We present a study of the $\rm{[OIII]\lambda\,5007\,\AA}$ line profile in a sub-sample of 8 active galactic nuclei (AGN) and 6 non-AGN in the optically-selected green valley at $\rm{z\,<\,0.5}$ using long-slit spectroscopic observations with the 11\,m Southern African Large Telescope. Gaussian decomposition of the line profile was performed to study its different components.
We observe that the AGN profile is more complex than the non-AGN one. In particular, in most AGN (5/8) we detect a blue wing of the line.
We derive the FWHM velocities of the wing and systemic component, and find that AGN show higher FWHM velocity than non-AGN in their core component. We also find that the AGN show blue wings with a median velocity width of approximately 600 $\rm{km\,s^{-1}}$, and a velocity offset from the core component in the range -90 to -350 $\rm{km\,s^{-1}}$, in contrast to the non-AGN galaxies, where we do not detect blue wings in any of their $\rm{[OIII]\lambda\,5007\,\AA}$ line profiles. Using spatial information in our spectra, we show that at least three of the outflow candidate galaxies have centrally driven gas outflows extending across the whole galaxy. Moreover, these are also the galaxies which are located on the main sequence of star formation, raising the possibility that the AGN in our sample are influencing SF of their host galaxies (such as positive feedback). This is in agreement with our previous work where we studied SF, morphology, and stellar population properties of a sample of green valley AGN and non-AGN galaxies.
\end{abstract}


\section{Introduction} \label{sec:intro}

It is often assumed that the essential properties of supermassive black holes (SMBH) in active galactic nuclei (AGN) and the galaxies that host them are connected \citep[e.g.,][and references therein]{Kormendy2013}. These properties include the relationship between black hole mass and stellar velocity dispersion, bulge mass, light concentration, bulge luminosity, etc. \citep[e.g.,][]{Povic2009, Povic2009b, Beifiori2012, Davis2019, DeNicola2019, Gaspari2019, Shankar2019, Caglar2020, Ding2020, Marsden2020, Bennert2021, Luo2021, Jelena2022, Matzko2022}. These relationships might be caused by an AGN feedback mechanism, where energy is transmitted from the nucleus to the host galaxy, which then slows the galaxy growth and quench star formation. However, AGN feedback may not be entirely responsible for these links \citep[e.g.,][]{Zubovas2013, Barai2018, DiPompeo2018, Wang2018}.

The AGN feedback operates in two general ways: by injecting energy into the surrounding gas of the interstellar or intergalactic matter, and by halting the in-fall of gas onto galaxies. These can then lead to matter compression, gas heating, enrichment or depletion by clearing gas within the galaxy -- all processes which in turn can either enhance star formation \citep[positive feedback; e.g., ][]{Silk2013, Cresci2015, Cresci2018, Shin2019, Perna2020}, or quench star formation \citep[negative feedback; e.g.,][]{Carniani2016, Karouzos2016, Kalfountzou2017, Bae2017, Cresci2018, Shin2019, Perna2020}. 

In recent years, significant effort has been dedicated to searching for observational signatures of AGN feedback using tracers of neutral atomic, molecular, and ionised gas, leading to the detection of outflows in galaxies  \citep[e.g.,][]{Hainline2013, Rupke2013, Harrison2014, Carniani2016, Fiore2017, Rupke2017, Liu2020, Lutz2020, Jarvis2020, Spilker2020, Spilker2020b,  Perna2020, Veilleux2017, Veilleux2020, AlYazeedi2021, BewketuBelete2021,  Scholtz2021, Paillalef2021}. 

One of the common methods to trace outflows uses forbidden emission lines to diagnose the dynamic state of the ionised gas in the AGN host galaxy \citep[e.g.,][]{Bae2017, Woo2020, Schmidt2021, Zhang2021}. In particular, AGN-driven outflows have been studied using the velocity measurements of the high-ionisation line $\rm{[OIII]\,\lambda5007\,\AA}$.\,The kinematics of $\rm{[OIII]}$ have typically been constrained by two measurements: a principal `core' component, with a velocity close to the systemic redshift of the host galaxy, and an additional broad `wing' component, which is asymmetric and usually stretches blue-ward \citep[e.g.,][]{Cresci2015,Kakkad2016, Kakkad2020,Sexton2021, Jelena2022}. The velocity-width of the core component is affected by the gravitational potential of the host galaxy and the central SMBH. But when the velocity-width of the blue wing is observed to be too broad, the gas cannot be in dynamical equilibrium with the host galaxy   \citep[e.g.,][]{Liu2013, Harrison2014, McElroy2015, Sun2017}, and commonly the blueshifted $\rm{[OIII]\,\lambda5007\,\AA}$ emission indicates gas outflows. Outflows of gas on the "other" side of the galaxy, which would be detected as redshifted, may be obscured due to dust, leaving only the blueshifted, near side of the outflow, observed.  Hence, the line profile, i.e. the line's width and (a)symmetry contains information about the dynamical state of the gas in the galaxy, and potentially its connection to the properties of the SMBH and the host galaxy $\rm{[OIII]\,\lambda5007\,\AA}$ emission profile \citep[e.g.,][]{Shen2014, Hryniewicz2022}.

Several studies have shown a correlation of outflow velocities detected in  $\rm{[OIII]\,\lambda5007\,\AA}$ with AGN luminosity in the optical, infrared (IR), and X-rays\,\citep[e.g.,][]{Zhang2011, Zakamska2014, Zakamska2016, Fischer2017, Perna2017, DiPompeo2018, Singha2022}, suggesting that radiation pressure is driving the winds. It has also been reported that outflow velocities have a relationship with other AGN properties such as black hole mass \citep[e.g.,][]{Rupke2017, Behroozi2019, Terrazas2020, Schmidt2021}, accretion rate \citep[e.g.,][]{Greene2005, Bian2006, Woo2016}, star formation \citep[e.g.,][]{Harrison2017, Fluetsch2019, Scholtz2020, Woo2020, Luo2021, Kim2022, Mulcahey2022}, although, in some cases, the correlation is weak at best \citep[e.g.,][]{Zhang2011, Peng2014}.

In \cite{Mahoro2017} we found, in a sample of 103 far-infrared (FIR) detected galaxies, that on average, star formation rates were higher in the (X-ray detected) AGN in the sample, than in the mass-matched non-AGN galaxies.  We also found that green valley X-ray detected AGN with FIR emission still have very active star formation rates, being located either on or above the main sequence (MS) of star-forming galaxies.\,We also found that they did not show signs of star formation quenching, but rather signs of its enhancement.\,This has been observed independently when examining their morphology \citep{Mahoro2019}. In addition, we did not find older stellar populations in AGN hosts, when studying stellar ages and populations of a subsample of AGN and non-AGN \citep{Mahoro2022},\,nor signs of star formation quenching, as suggested previously in X-ray and optical studies\,\citep[e.g.,][]{Nandra2007, Povic2012, Ellison2016, Leslie2016}. Therefore, our results may suggest that if AGN feedback influences the star formation in green valley galaxies with X-ray detected AGN and FIR emission, it is {\em positive} AGN feedback rather than negative feedback. 

In this work, we go a step further in studying in detail the possibility of detecting signs of positive AGN feedback in our sample, by analysing the $\rm{[OIII]\lambda5007\,\AA}$ emission line profiles in a sub-sample of FIR AGN and non-AGN green valley galaxies, using our own observations from the 11m Southern African Large Telescope (SALT).

The paper is organised as follows:\,a description of the sample and observations is given in Section 2.\,Section 3 presents the spectroscopic fitting procedure, while the main results are presented in Section 4.\,Discussion and summary are given in Sections 5 and 6. Throughout, we assume a flat universe with $\Omega_{m}=0.3,\,\Omega_{\Lambda}=0.7$ and $H_{0}=70$\,km\,s$^{-1}$\,Mpc$ ^{-1}$ cosmology. In addition, we assume \cite{Salpeter1955} initial mass function (IMF). All magnitudes are in the AB system, and the stellar masses are given in units of solar masses (M$_{\odot}$).

\section{Data and the Sample}

The sample studied in this work was taken from a much larger sample of optically-selected green valley galaxies using $\rm{U-B}$ rest-frame colour criteria of
$\rm{0.8\leq U-B\leq 1.2}$. Furthermore, the obtained optically-selected green valley sample was then cross-matched with \textit{Herschel}/PACS FIR data \citep[for a more detailed description see][]{Mahoro2017}. Thus, the sample consists of optically-selected green valley galaxies with FIR detections.

The data to study our sample were obtained using the 11-m class SALT \citep{Buckley2006}, and its Robert Stobie Spectrograph (RSS) \citep{Burgh2003, Kobulnicky2003}. The RSS is a versatile instrument providing several observing modes and is designed to work in a wavelength range of $\rm{3200-9000\,\AA}$.

For our study of the $\rm{[OIII]\lambda5007\,\AA}$ line profile, we first selected the AGN and non-AGN green valley galaxies from \cite{Mahoro2017} observable with SALT, and within the magnitude limit of Vmag $\rm{\leq}$ 21.5, in order to have reasonable exposure times. This resulted in 11 AGN and 669 non-AGN galaxies. Beyond approximately 7500 $\rm{\AA}$, there are strong sky emission lines which affect the spectral frames and therefore the accuracy of data reduction.\,Since our galaxies are fairly faint, we further selected only  those with redshifts $\rm{<}$ 0.5, to avoid the observed emission line falling in this region of the spectral frames.\,We thus finally obtained nine AGN galaxies.\,From the larger number of non-AGN that satisfied those criteria, we selected a smaller {\em matching} sample by selecting the one that was brightest, and in particular {\em closest} to each selected AGN galaxy in terms of its location in a stellar mass versus star formation rate diagram.  We thus obtained a final sample of nine non-AGN galaxies also. \autoref{SALT_main} shows both the AGN and non-AGN sub-samples on the star formation rate vs. stellar mass relation diagram. We used the results by \cite{Elbaz2011}, which were obtained from galaxies detected by Herschel, to represent the MS of star-forming galaxies. This MS is based on a Salpeter IMF \citep{Salpeter1955}.

Note also, that using low-resolution Sloan Digital Sky Surveys fourth phase\footnote{\url{https://www.sdss.org}}\citep[SDSS-IV,][]{Blanton2017} spectra and zCOSMOS survey spectra \citep{Lilly2009}, we checked that all our AGN galaxies are type-2 AGN.
\textbf{
\begin{figure}[!ht]
\centering
\includegraphics[width=\columnwidth]{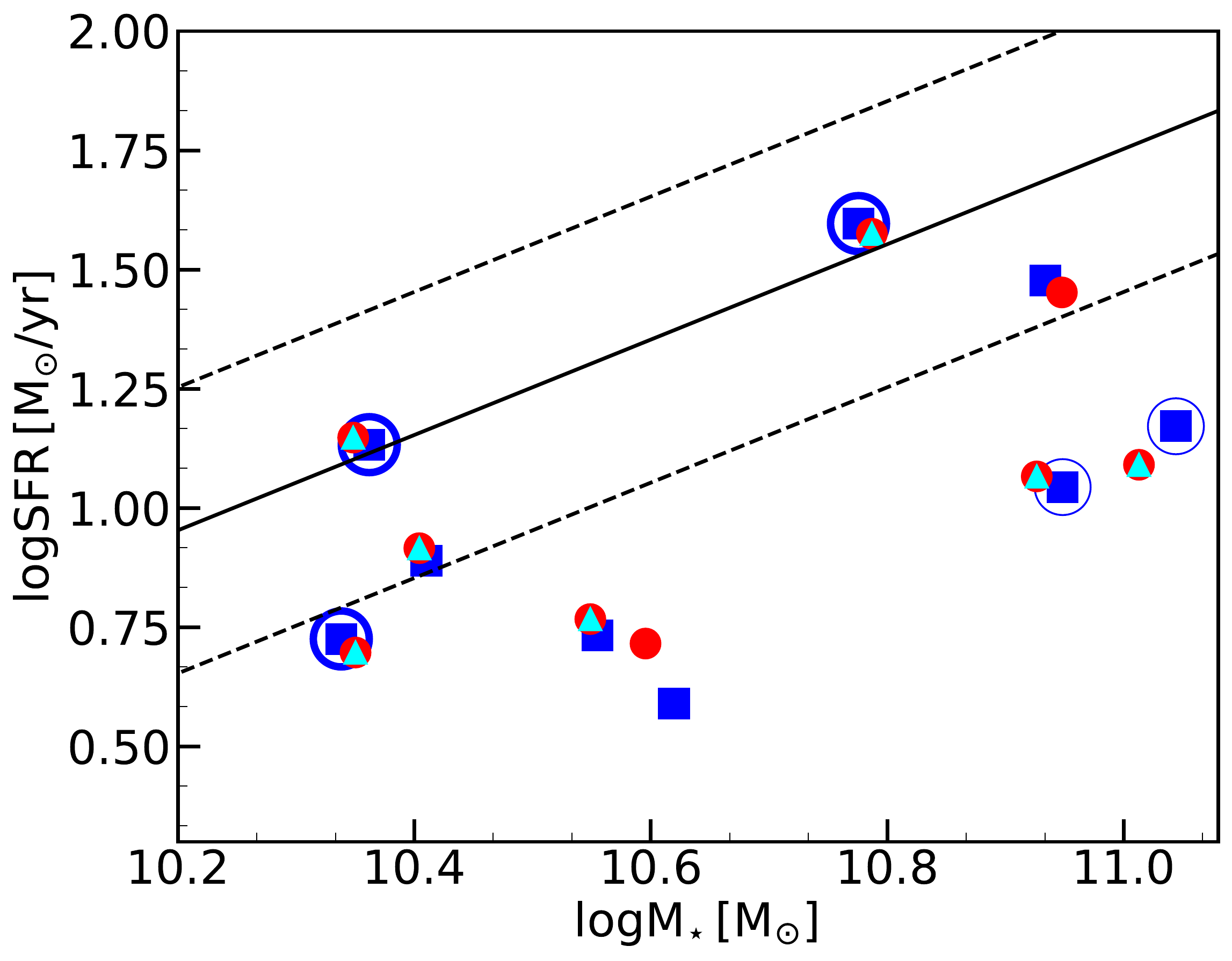}
\caption{Star formation rate vs. stellar mass. AGN are represented by blue squares, while the red circles represent seven non-AGN. The cyan triangles are those non-AGN that finally were observed, while all the AGN were observed. The solid black line shows the \protect\cite{Elbaz2011} fit for MS, while the dashed lines represent the MS width of $\rm{\pm}$ 0.3 dex. Open blue circles show the cases with detected blue wings in their emission line profiles (see \protect\autoref{sec:results}), where the three sources with unambiguous gas flow interpretations are marked with a thick blue circle (see \protect\autoref{sect:discussion}).}
\label{SALT_main}
T\end{figure}}

\subsection{Observations and data reduction}\label{data}

In this section, we describe all observations done and the data reduction process. The observations were planned and defined using the RSS simulator tool\footnote{\url{https://astronomers.salt.ac.za/software/\#RSS}}. In total, 16 sources were observed with the SALT/RSS, as part of the programs 2017-2-MLT-003 (PI:\,Povi\'c M.),\,2020-1-DDT-001 and 2020-2-SCI-021\,(PI:\,Mahoro A). \autoref{tab:obslog} summarises the observational information, including average seeing and exposure times for all 16 observed AGN and non-AGN. The [OIII]$\rm{\lambda5007\,\AA}$ emission line was detected in 14 of the target spectra (8 AGN and 6 non-AGN).

All RSS observations were taken in the long-slit mode using a 1.25 arcsec wide slit with PG1800 or PG2300  gratings (column 8 in \autoref{tab:obslog}). In most cases, the slit was placed along the galaxy minor axis. Figures with the slit position are shown in \autoref{slit_images}, while the position angle is given in \autoref{tab:obslog} (column 9). RSS uses an array of three CCDs of $\rm{2048\,\times\,4096}$ pixels, and the grating angle in the instrument was set in a way to avoid the [OIII]$\rm{\lambda5007\,\AA}$ emission line falling in the resulting two 15 pixels size gaps. Interpolated pixel values were used to fill in the two CCD gaps.\,The RSS spatial pixel scale is 0.1267 arcsec, and the effective field of view is 8 arcmin along the slit.\,We used a binning factor of 2 for a final spatial sampling of 0.253 arcsec pixel$^{-1}$.\,It is known that on RSS, using spectral flats, the improvement of rms noise in data below 7000\AA\, is less than $\rm{<}$ 5 \%, i.e. quite insignificant, and we thus chose to use the time for extra exposure time instead\footnote{\url{https://pysalt.salt.ac.za/proposal_calls/current/ProposalCall.html\#h.sc2wkzpxrdkc}}.\,However, above 7500\AA\, there is significant fringing, which can only be corrected using the spectral flats, and hence for all setups extending to that wavelength range we took them. As our AGN are at a slightly higher redshift than non-AGN (see \autoref{tab:obslog}, column 3), and the observed wavelength can reach 7000\AA, these spectra were taken with spectral flat-fields, while the non-AGN were not.
\vspace{-4mm}

\begin{deluxetable*}{lccccccccccl}
\tablecolumns{10}
\tabletypesize{\footnotesize}
\tablecaption{Summary of the properties of our sources and the instrumental set-up for the observations.\label{tab:obslog}}
\tablehead{
\colhead{ID}            &
\colhead{RA(J2000)}     &
\colhead{Dec(J2000)}    &
\colhead{Redshift}      &
\colhead{V-band}        &
\colhead{Seeing}        &
\colhead{Total Exposure}&
\colhead{Grating}       &
\colhead{Pos. ang.}     &
\colhead{Galaxy Type}  \\
\colhead{}              &
\colhead{hours}         &
\colhead{degrees}       &
\colhead{}              &
\colhead{AB(mag)}       &
\colhead{(arcsec)}      &
\colhead{time (s)}      &
\colhead{}              &
\colhead{degrees}       &  
\colhead{}  
}
\startdata
COSMOS\_17759  & 09h 59m 07.66s  & +02$^{\rm{\circ}}$ 08$^{'}$ 20.80$^{''}$   &0.354   &20.50   & 1.35     & 3130          & PG2300  &  50.0     & AGN  \\
COSMOS\_29436  & 09h 58m 38.84s  & +02$^{\rm{\circ}}$ 23$^{'}$ 48.40$^{''}$   &0.355   &20.10   & 1.25     & 1862          & PG2300  &  83.0     & AGN  \\
COSMOS\_20881  & 10h 01m 40.41s  & +02$^{\rm{\circ}}$ 05$^{'}$ 06.70$^{''}$   &0.424   &21.19   & 1.5      & 6405          & PG1800  &  330.0    & AGN  \\
COSMOS\_8450   & 09h 59m 26.89s  & +01$^{\rm{\circ}}$ 53$^{'}$ 41.20$^{''}$   &0.443   &20.93   & 1.8      & 1500          & PG1800  &  15.5     & AGN  \\
COSMOS\_11702  & 09h 59m 47.50s  & +01$^{\rm{\circ}}$ 38$^{'}$ 52.60$^{''}$   &0.283   &21.29   & 1.2      & 6405          & PG2300  &  83.0     & AGN  \\
COSMOS\_28917  & 09h 58m 40.65s  & +02$^{\rm{\circ}}$ 04$^{'}$ 26.60$^{''}$   &0.339   &20.40   & 1.25     & 1365          & PG2300  &  4.0      & AGN  \\
COSMOS\_9239   & 10h 02m 36.52s  & +02$^{\rm{\circ}}$ 02$^{'}$ 17.60$^{''}$   &0.368   &20.66   & 1.7      & 4270          & PG2300  &  293.0    & AGN  \\
COSMOS\_30402  & 09h 58m 40.32s  & +02$^{\rm{\circ}}$ 08$^{'}$ 07.00$^{''}$   &0.338   &20.39   & 1.6      & 4270          & PG2300  &  44.0     & AGN  \\ 
COSMOS\_12899$^{*}$  & 09h 59m 09.57s  & +02$^{\rm{\circ}}$ 19$^{'}$ 16.30$^{''}$   &0.383   &21.29   & 1.5      & 5847          & PG2300  &  343.0    & AGN  \\
COSMOS\_6026   & 10h 02m 41.69s  & +02$^{\rm{\circ}}$ 13$^{'}$ 34.70$^{''}$   &0.214   &19.93   & 1.5      & 2970          & PG2300  &  320.0    & Non-AGN \\
COSMOS\_7952   & 09h 58m 02.19s  & +02$^{\rm{\circ}}$ 48$^{'}$ 12.70$^{''}$   &0.249   &19.96   & 1.1      & 2970          & PG2300  &  3.5      & Non-AGN \\
COSMOS\_19854  & 10h 00m 47.40s  & +02$^{\rm{\circ}}$ 15$^{'}$ 33.60$^{''}$   &0.214   &19.91   & 1.7      & 2235          & PG2300  &  90.0     & Non-AGN \\  
COSMOS\_21658  & 09h 58m 21.48s  & +02$^{\rm{\circ}}$ 25$^{'}$ 18.60$^{''}$   &0.187   &19.94   & 1.3      & 3148          & PG2300  &  29.0     & Non-AGN \\  
COSMOS\_21854  & 09h 59m 04.99s  & +02$^{\rm{\circ}}$ 35$^{'}$ 06.60$^{''}$   &0.219   &19.95   & 1.5      & 2670          & PG2300  &  289.0    & Non-AGN \\  
COSMOS\_22743  & 10h 00m 05.49s  & +02$^{\rm{\circ}}$ 40$^{'}$ 02.50$^{''}$   &0.218   &19.96   & 1.5     & 2970          & PG2300  &  347.0    & Non-AGN \\ 
COSMOS\_14973$^{*}$  & 10h 02m 22.51s  & +01$^{\rm{\circ}}$ 52$^{'}$ 28.30$^{''}$   &0.252   &19.95   & 1.7      & 2870          & PG2300  &  66.0     & Non-AGN   
\enddata
\tablecomments{Sources observed with no detection of $\rm{[OIII]\lambda\,5007\,\AA}$ emission line are marked with '*'.\\
}
\end{deluxetable*}
\vspace{-4mm}
We made use of the SALT primary reduced product data sets, generated by the in-house pipeline called PySALT\footnote{\url{http://pysalt.salt.ac.za}} \citep[see][]{Crawford2010}, which mosaics the individual CCD data to a single FITS file, corrects for cross-talk effects, and performs bias and gain corrections.\,Further steps, including flat-field correction (for AGN only) were done by us with the help of the IRAF package\footnote{IRAF is distributed by the National Optical Astronomy Observatories, which are operated by the Association of Universities for Research in Astronomy, Inc., under cooperative agreement with the NSF.}. Two consecutive exposures were combined to increase the signal-to-noise ratio, and to remove the cosmic ray effects using the L.A.Cosmic task in IRAF \citep{vanDokkum2001}.\,Using Arc lamp spectra (Argon, Neon-Argon, and Xenon), we determined the dispersion solution using the noao.twodspec package.\,After wavelength calibration, the RMS error in the wavelength solution was $\rm{\sim\,0.02\,\AA}$. Flux calibration was performed using spectrophotometric standard stars (LTT4364, HILT600 and LTT3218).\,The standard stars were observed with the same settings as our respective observation blocks.
Note that in this work the flux calibration was done for the purpose of determining the {\em relative} spectral shape, as an absolute flux calibration is not required. 

\section{Spectral fitting}\label{fittingprocess}
The main aim of this paper is to analyse the [OIII]$\rm{\lambda5007\,\AA}$ emission line profile.\,Before [OIII]$\rm{\lambda5007\,\AA}$ line profile fitting, we did a cross-check with all spectra and applied a three-pixel and five-pixel boxcar smoothing separately to all our spectra.\,After visual checking, we do not see a significant change in the spectral feature to be analysed between the two cases, and in addition we made sure that the latter smoothing kernel corresponds to the spectral resolution of the instrumental setup. Thus, in order to maximise the signal-to-noise ratio in the data, we decided to use the five-pixel boxcar smoothed spectra for our analysis.

\begin{figure*}[!ht]
\centering
    \includegraphics[width=.23\linewidth,clip]{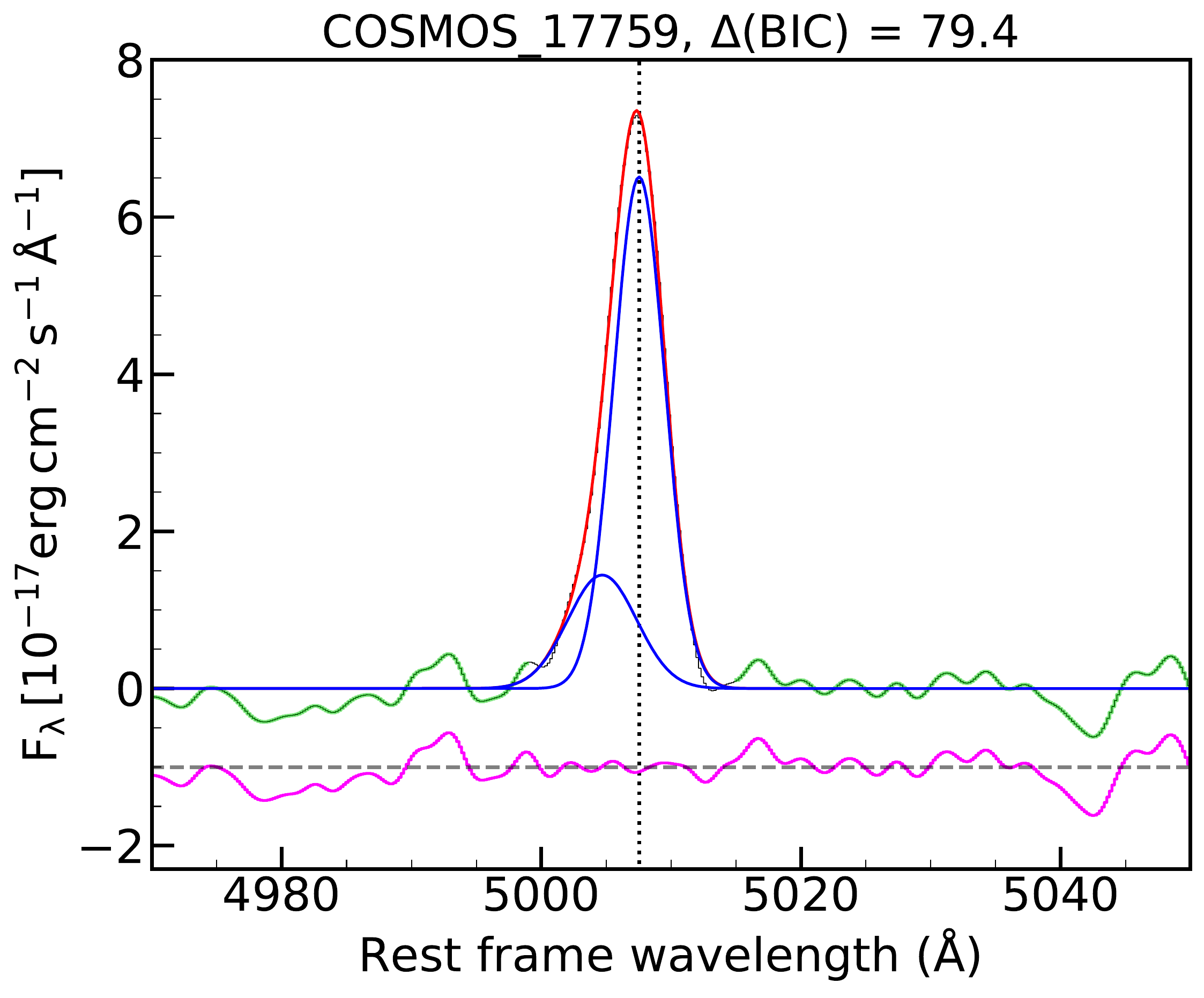}
    \includegraphics[width=.23\linewidth,clip]{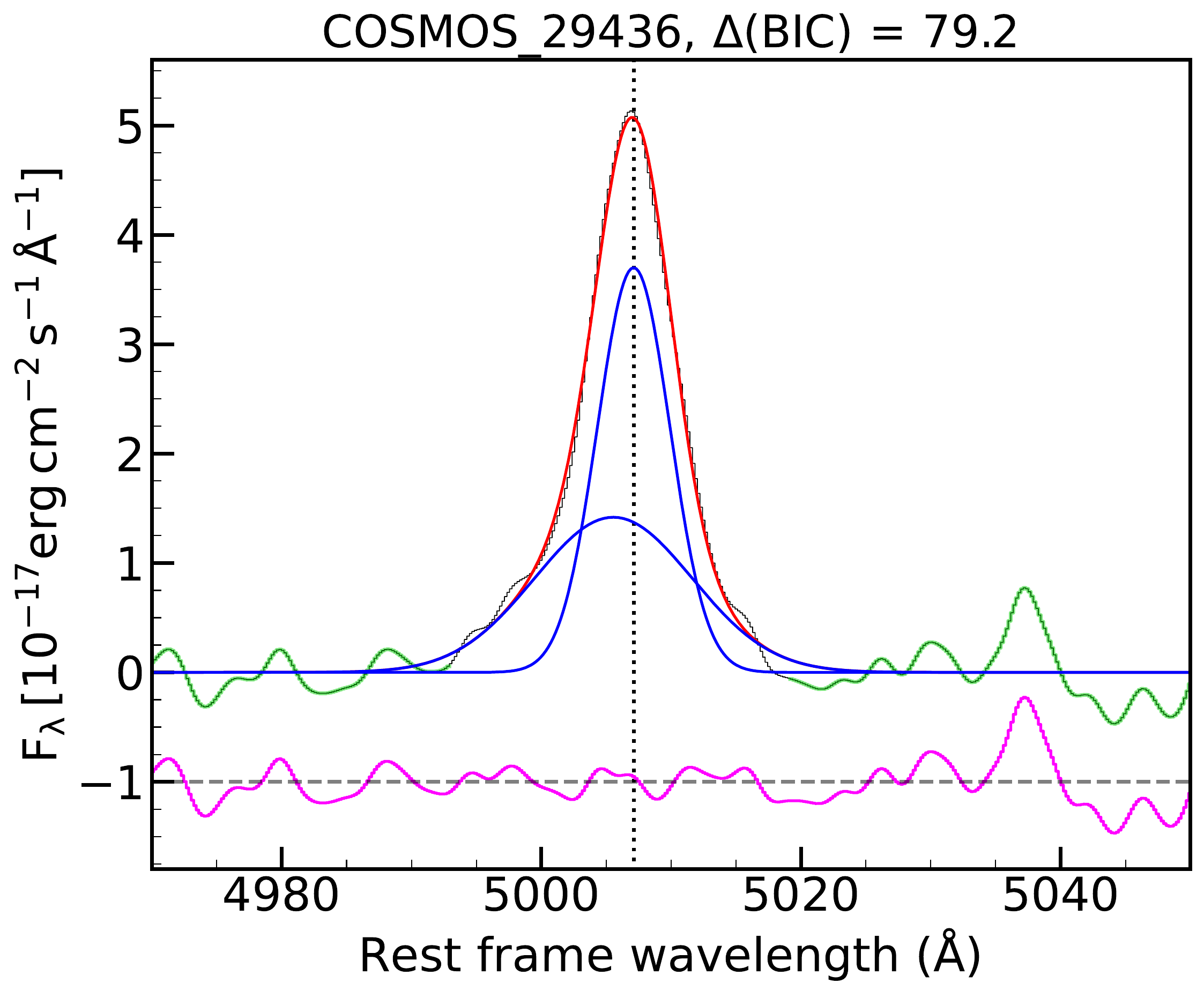}
    \includegraphics[width=.23\linewidth,clip]{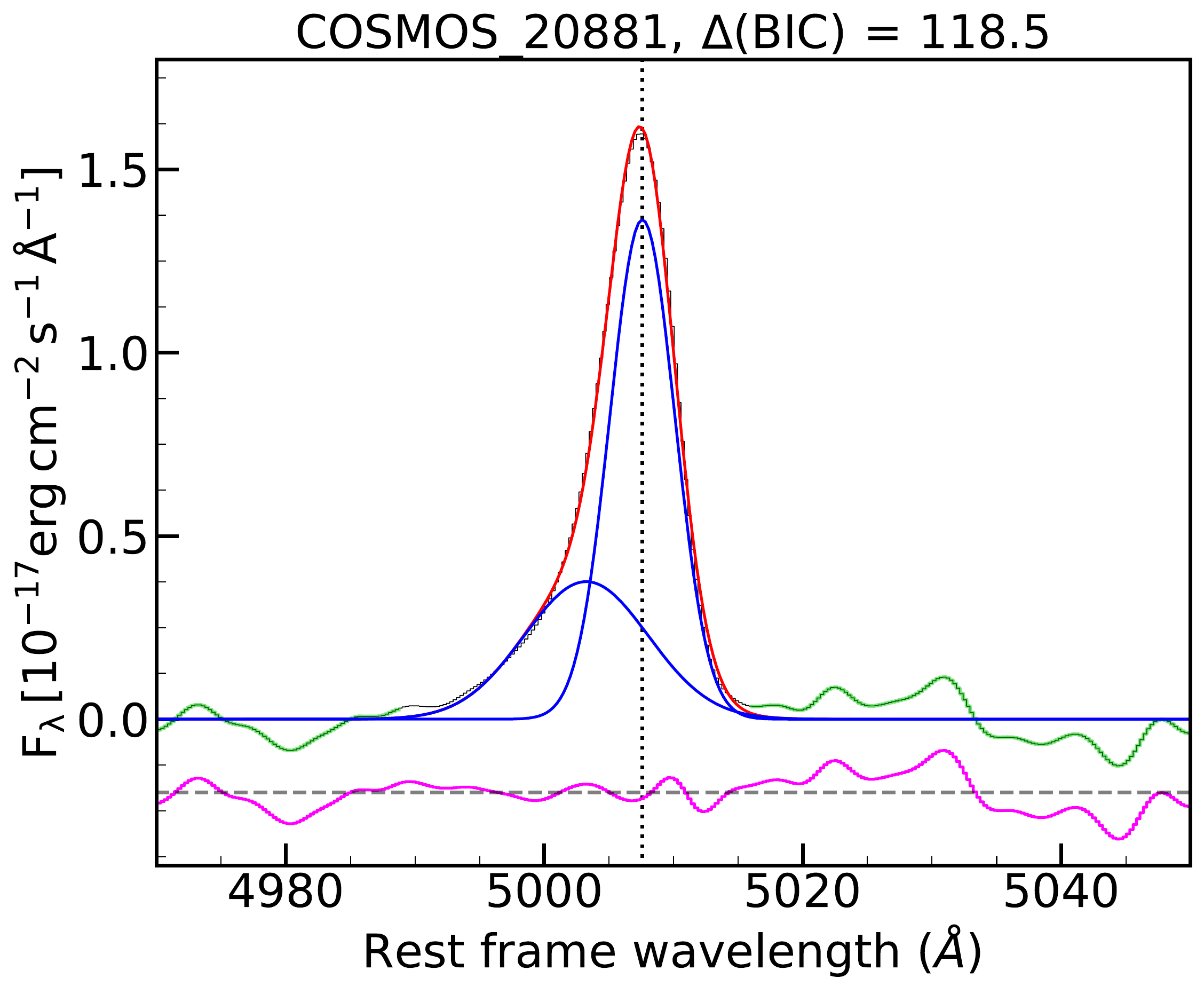}
    \includegraphics[width=.23\linewidth,clip]{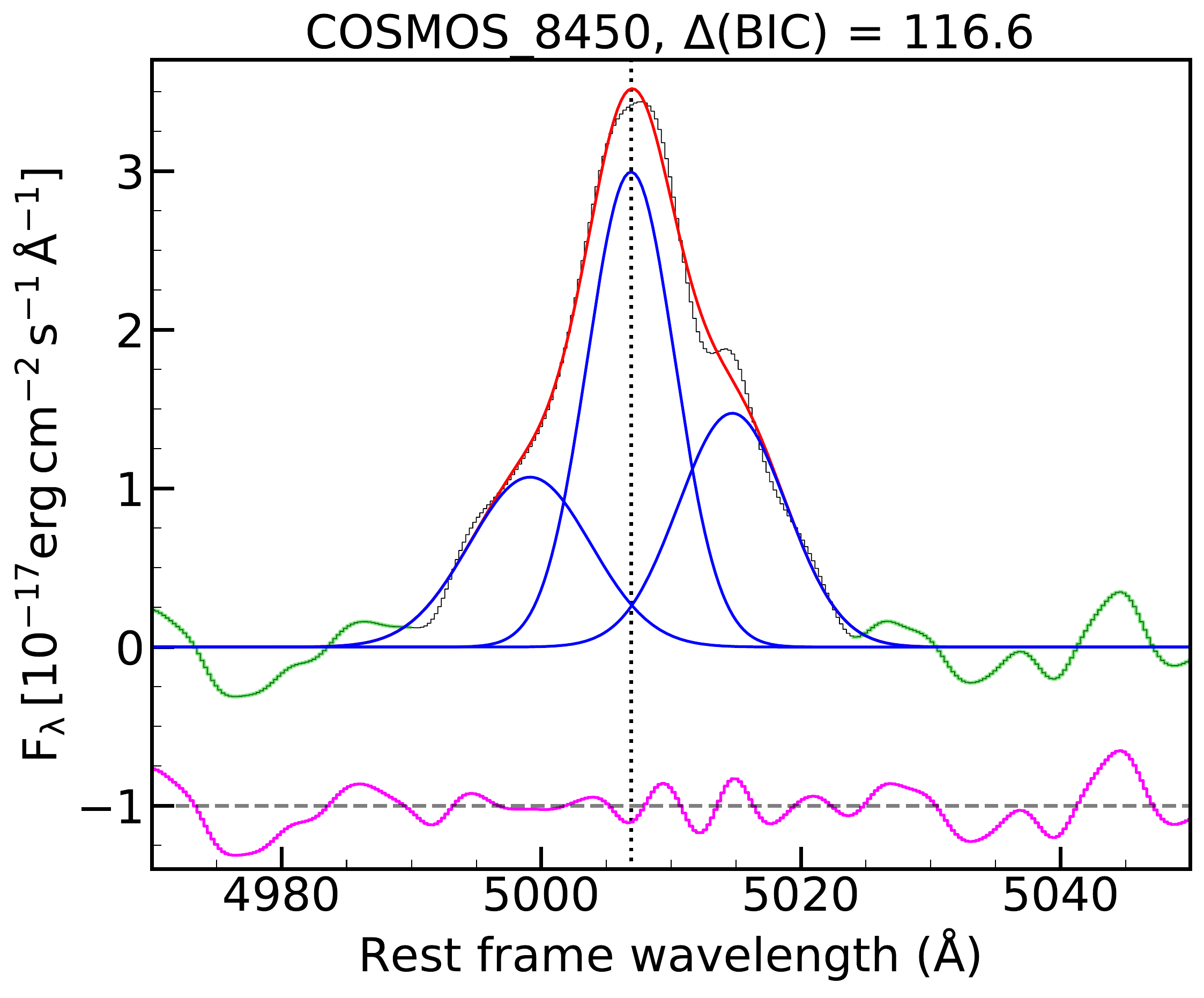} \\
    \includegraphics[width=.23\linewidth,clip]{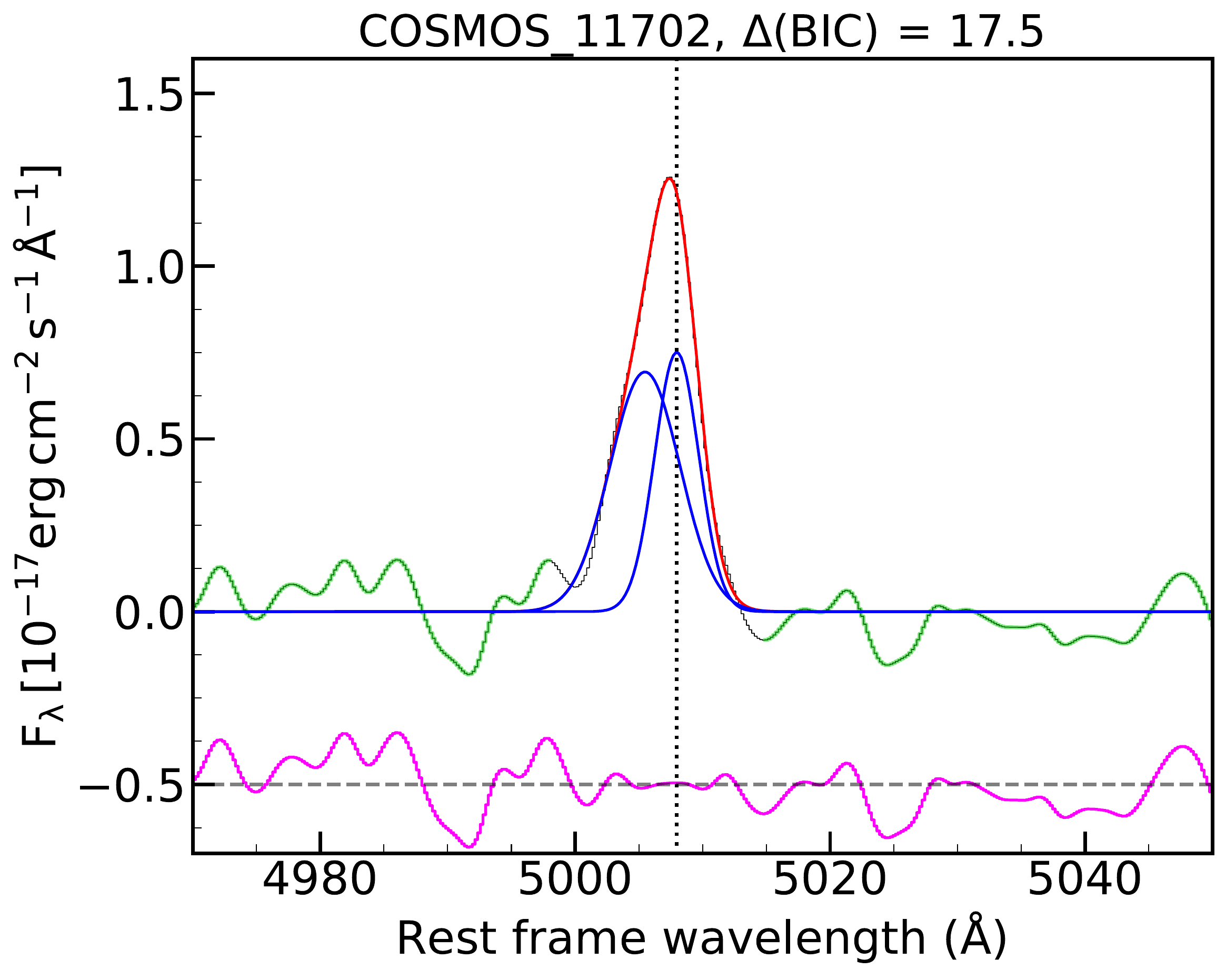}
    \includegraphics[width=.23\linewidth,clip]{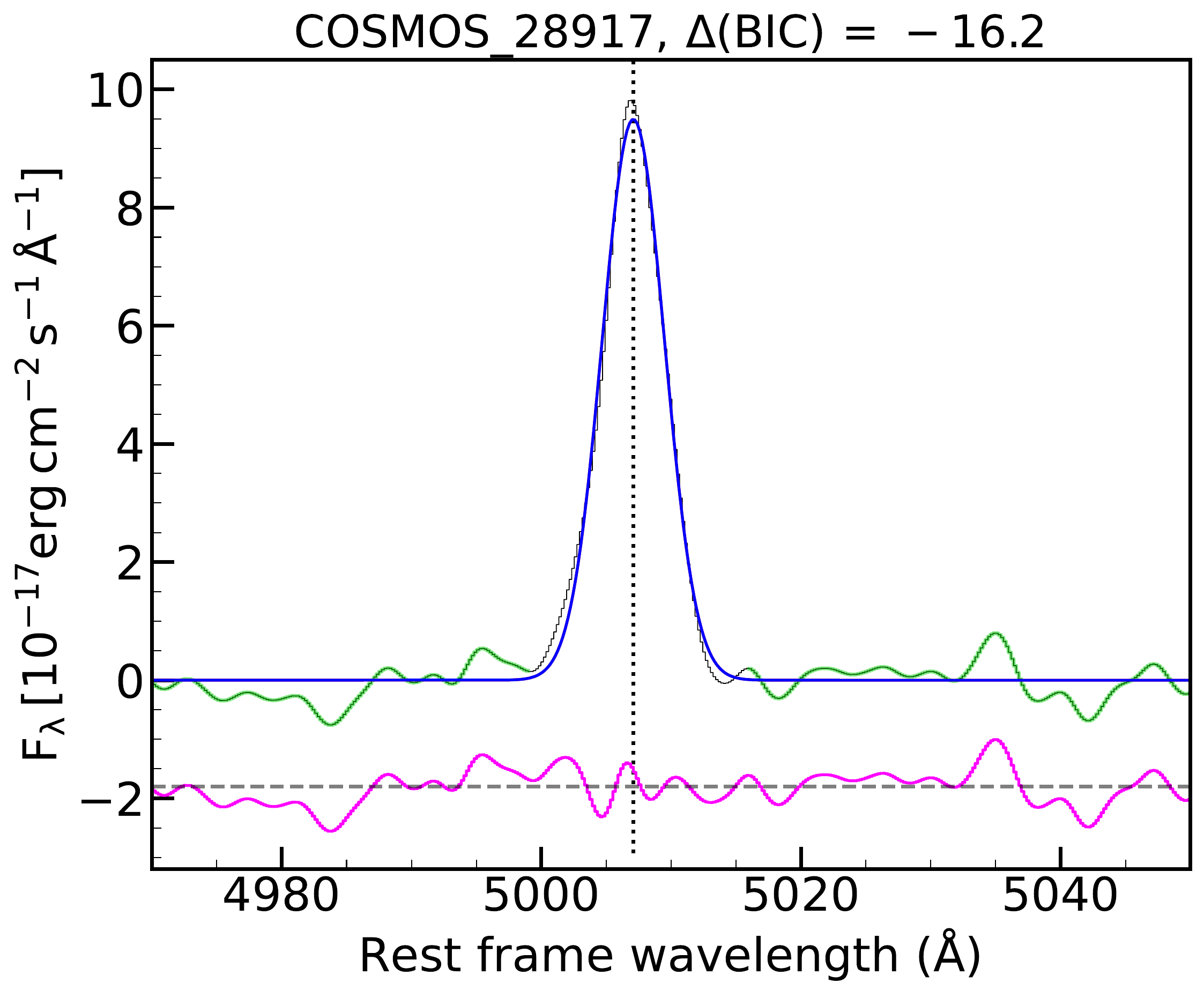}
    \includegraphics[width=.23\linewidth,clip]{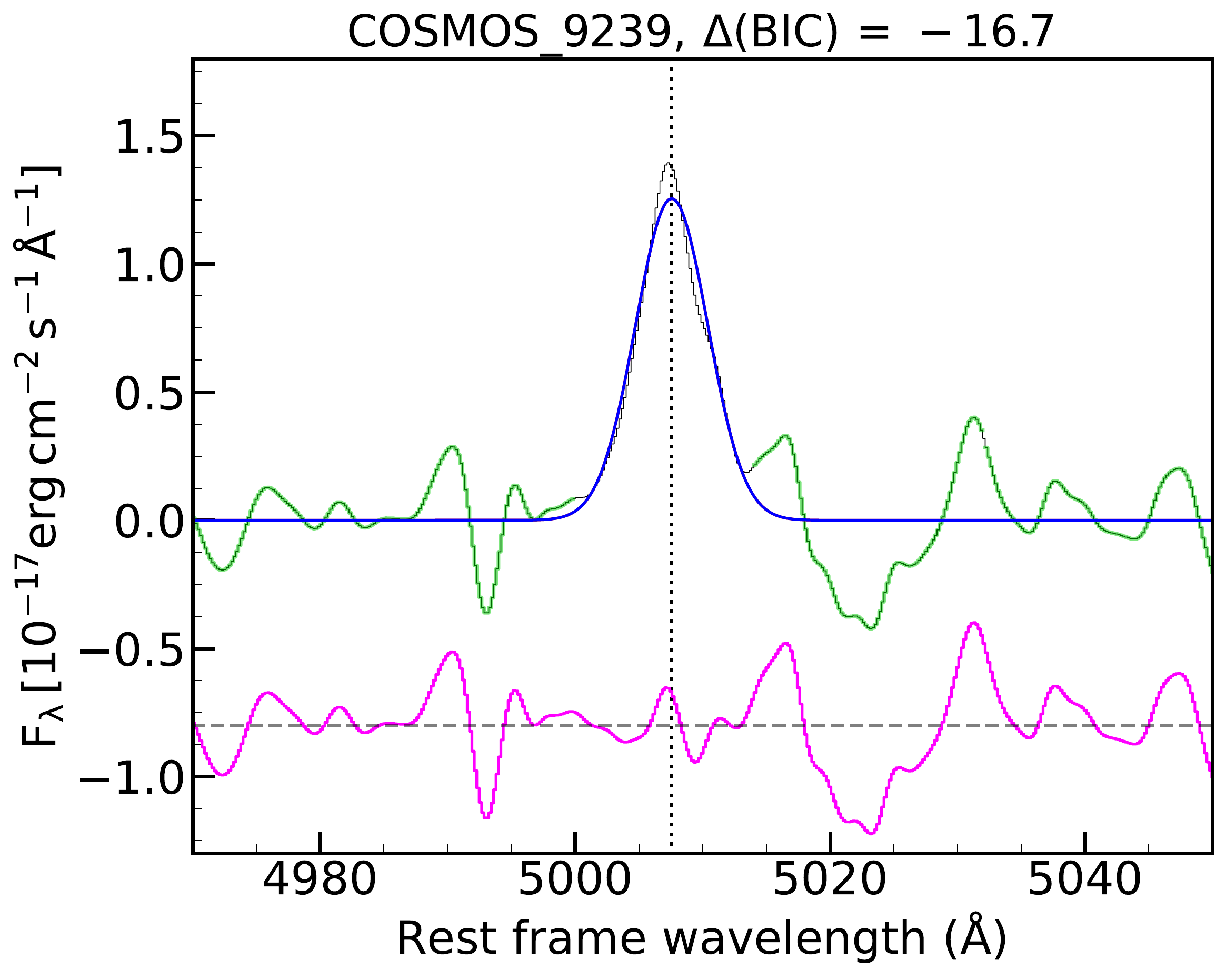}
    \includegraphics[width=.23\linewidth,clip]{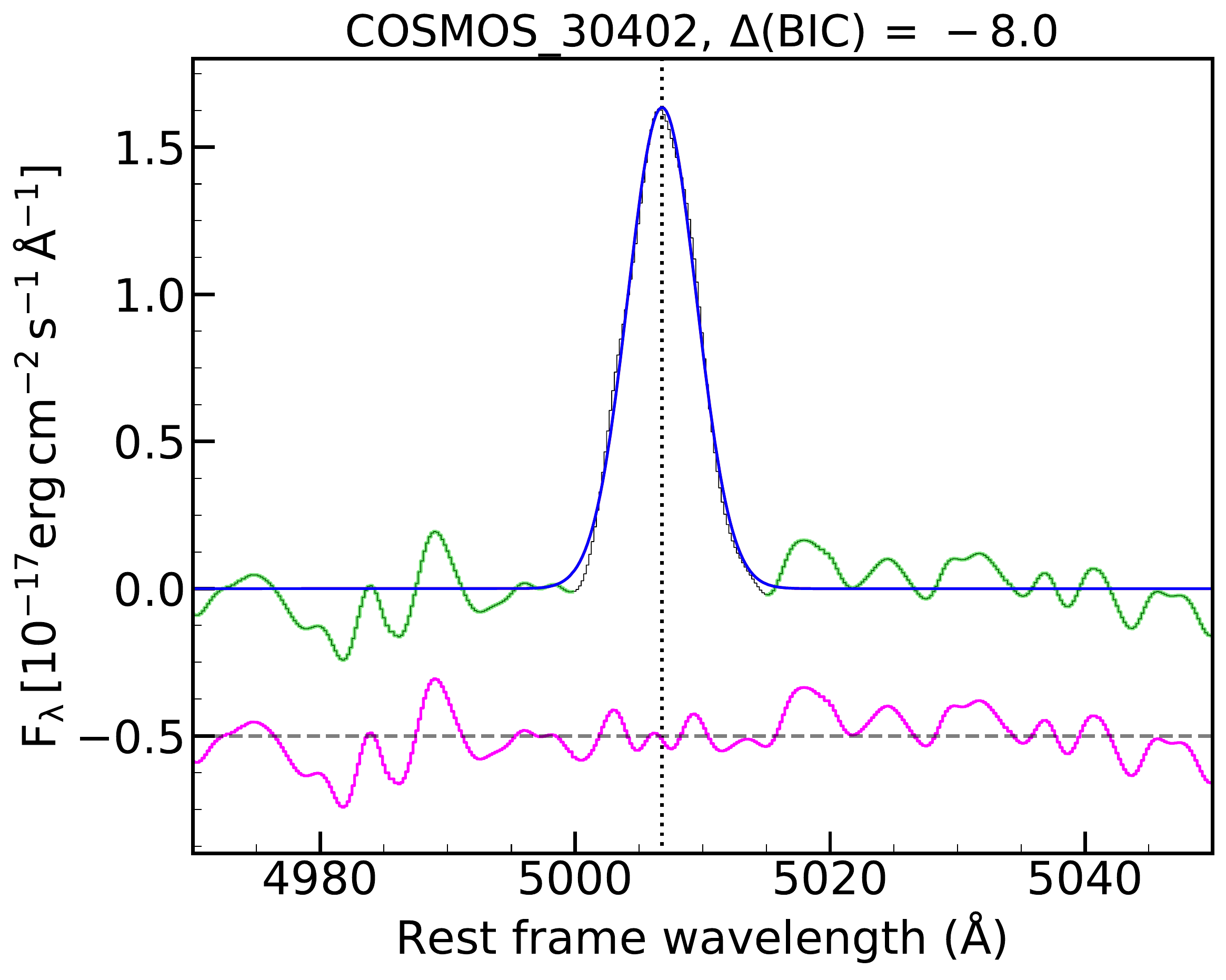}\\
    \includegraphics[width=.23\linewidth,clip]{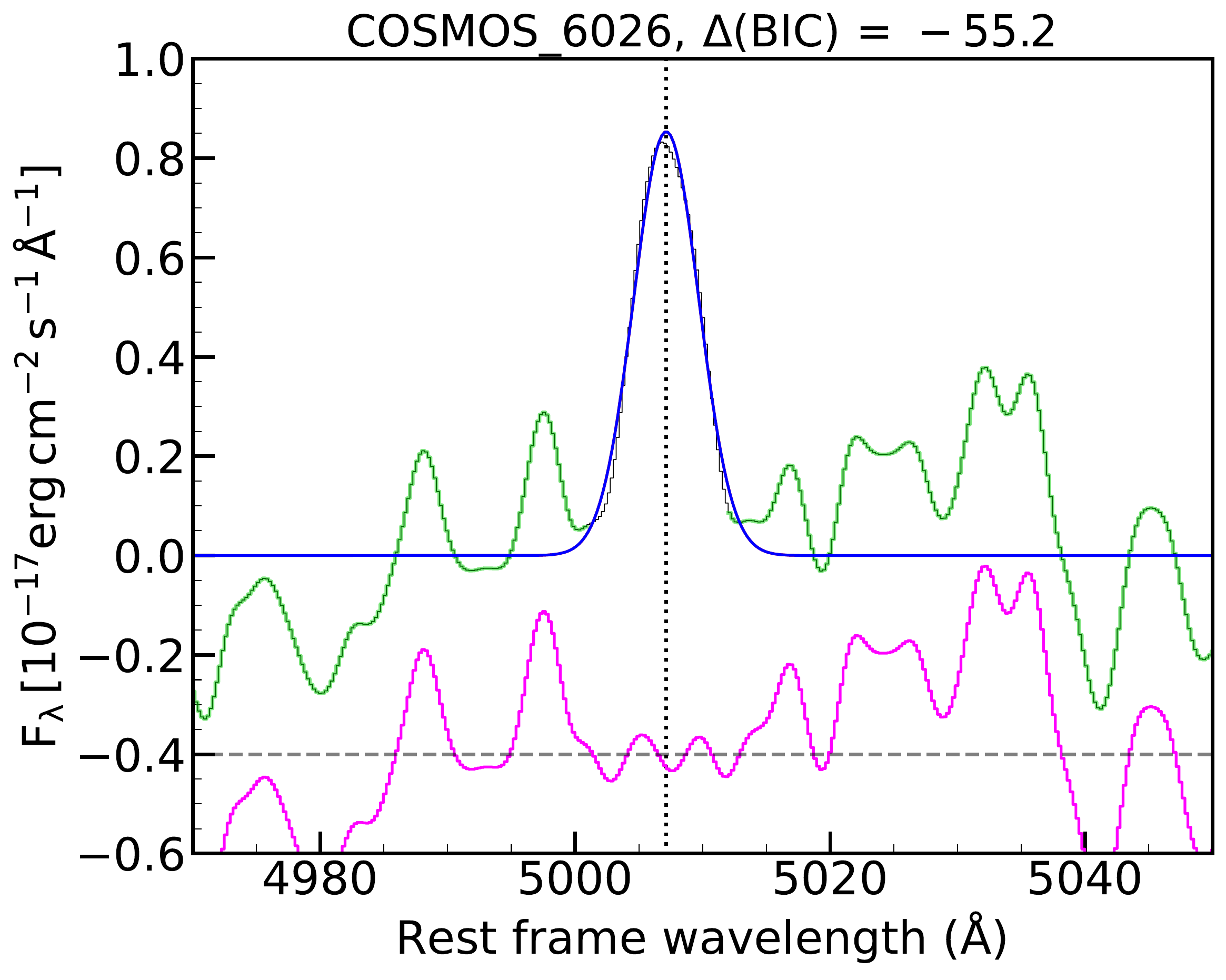}
    \includegraphics[width=.23\linewidth,clip]{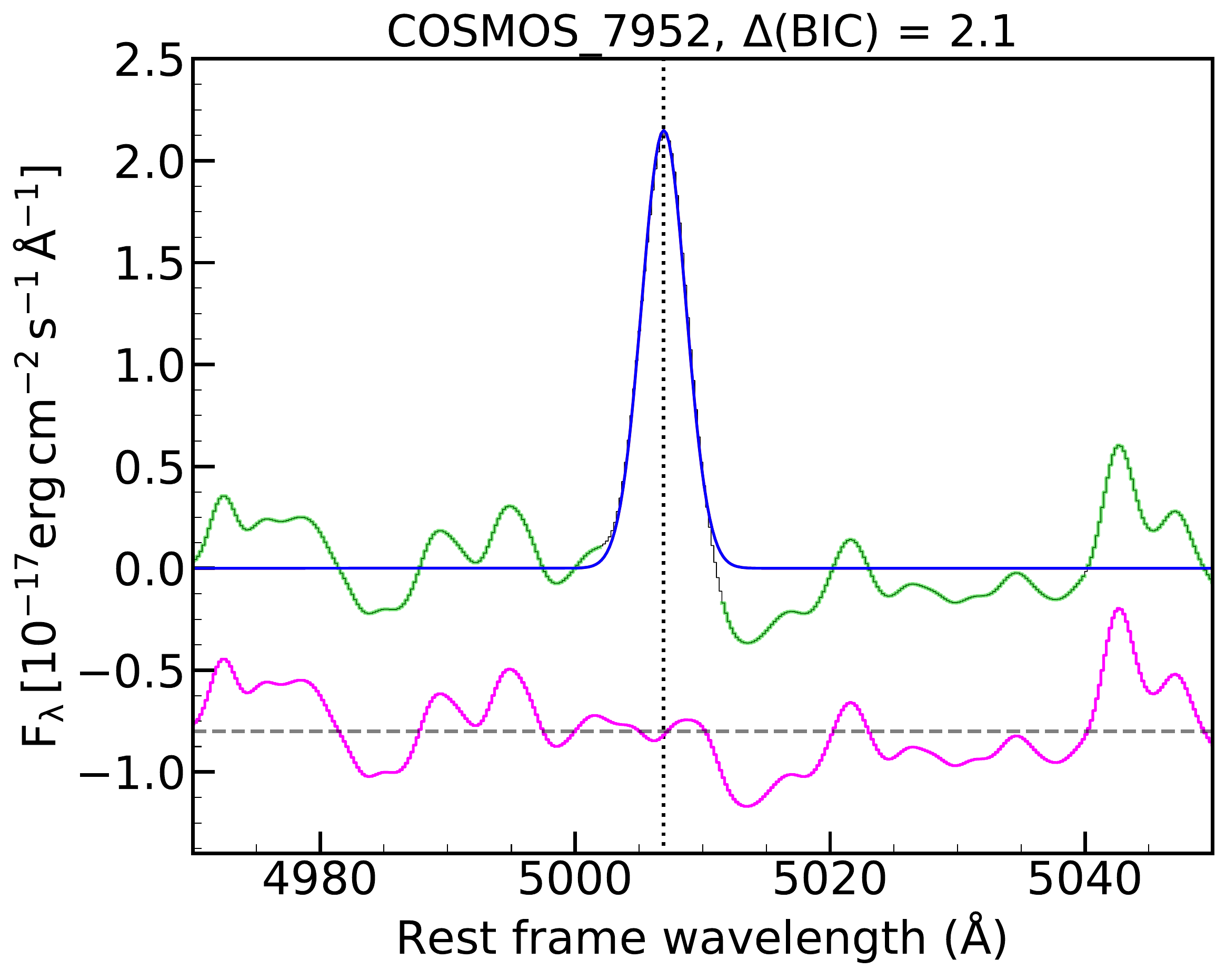}
    \includegraphics[width=.23\linewidth,clip]{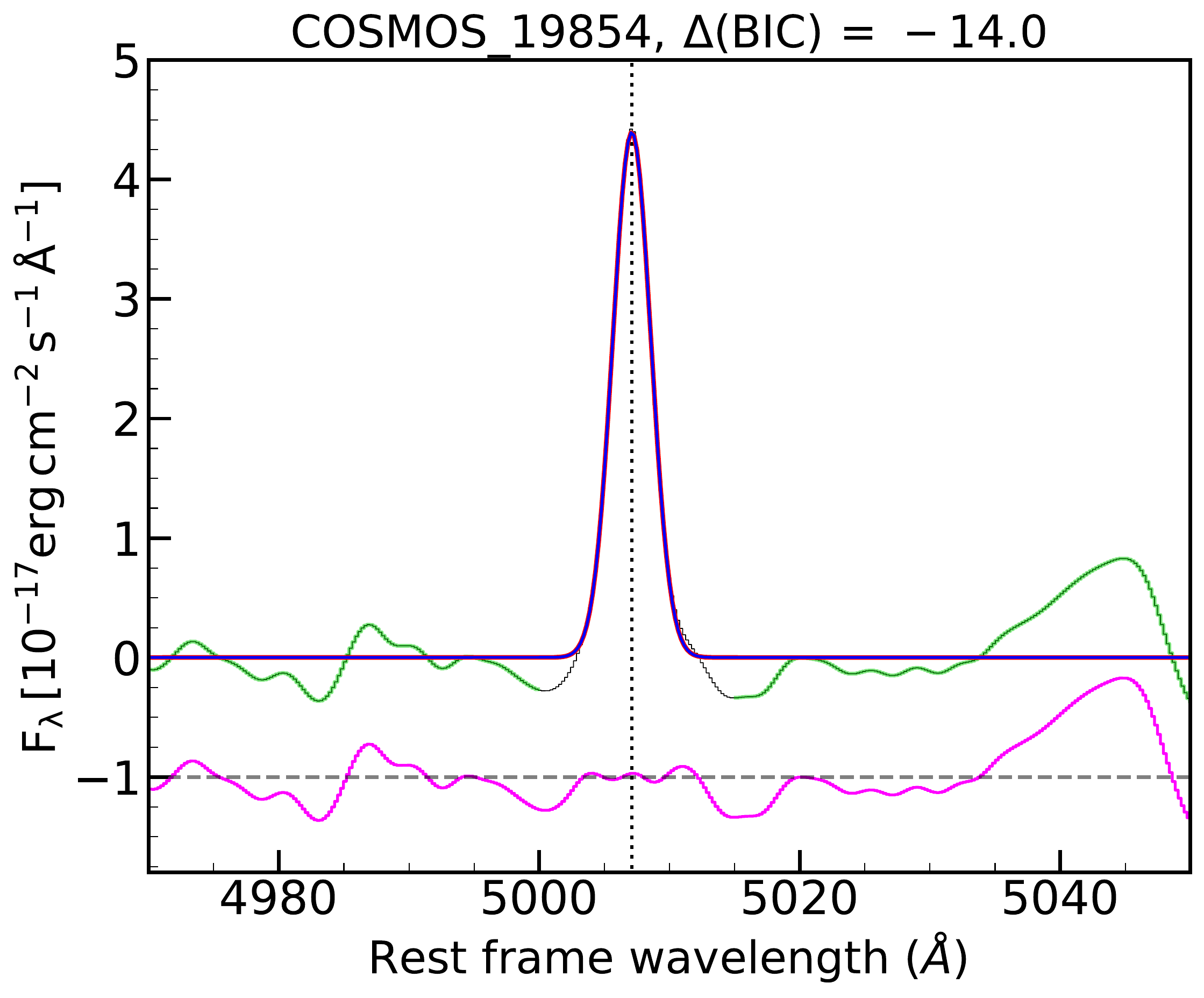}
    \includegraphics[width=.23\linewidth,clip]{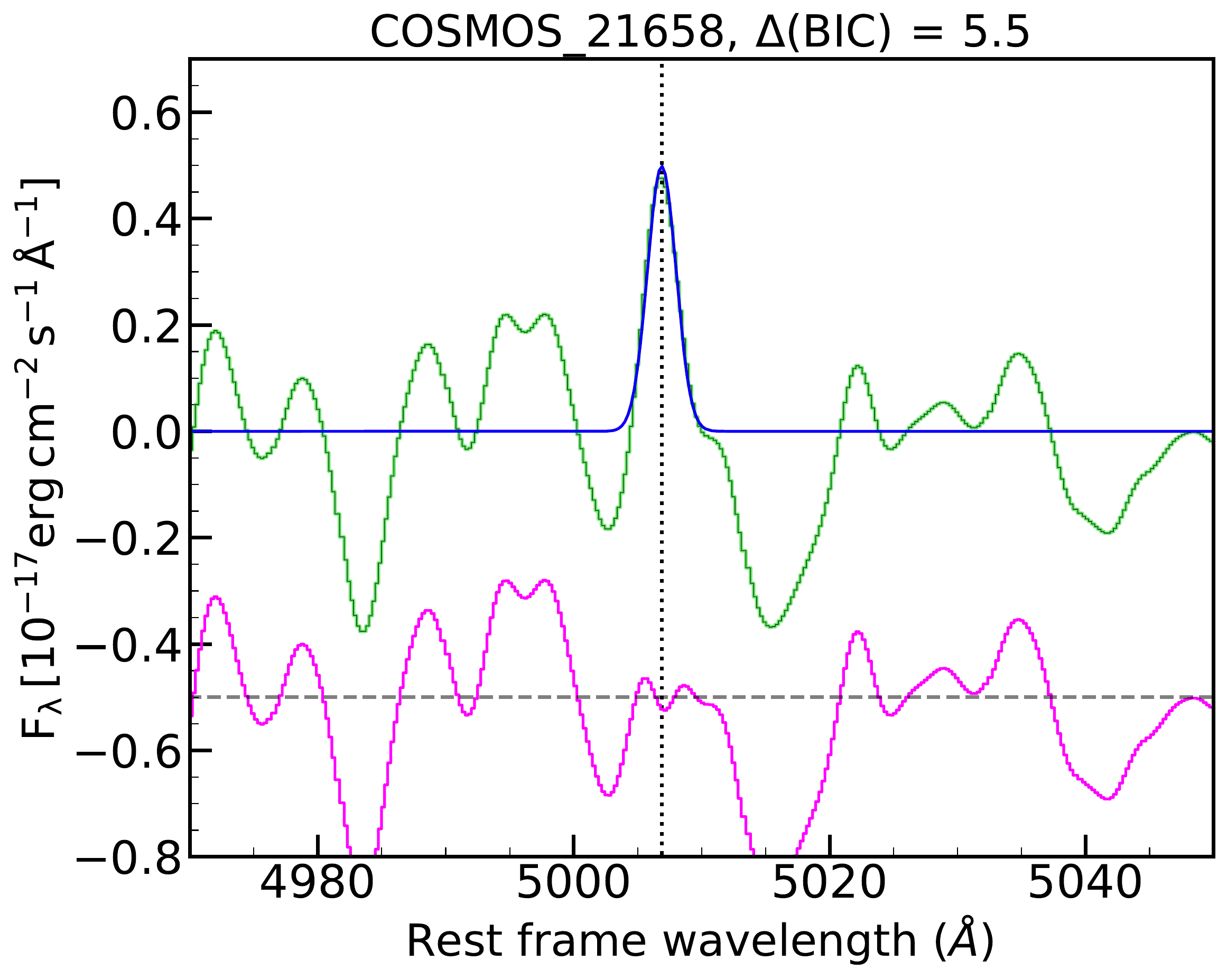}\\
    \includegraphics[width=.23\linewidth,clip]{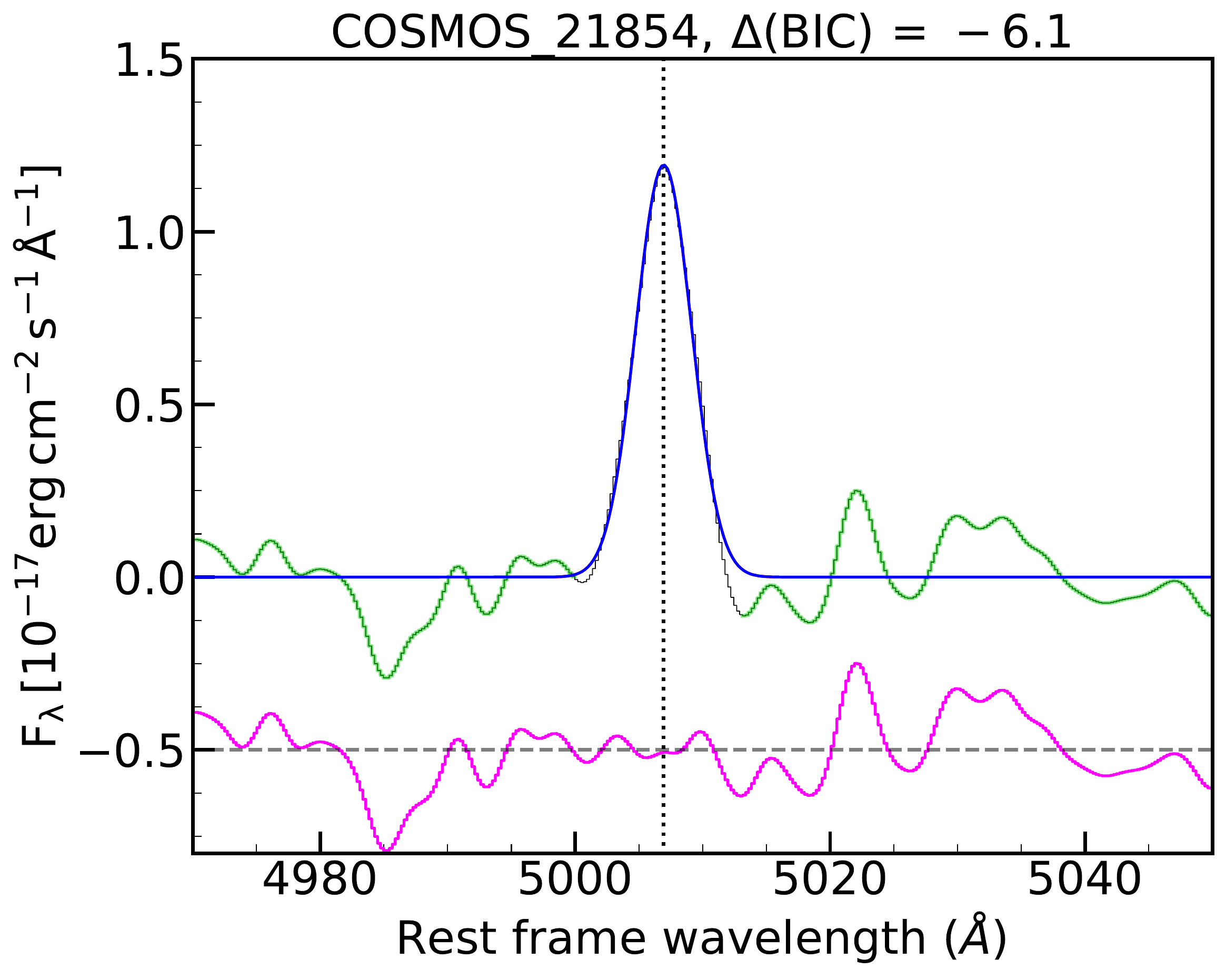}
    \includegraphics[width=.23\linewidth,clip]{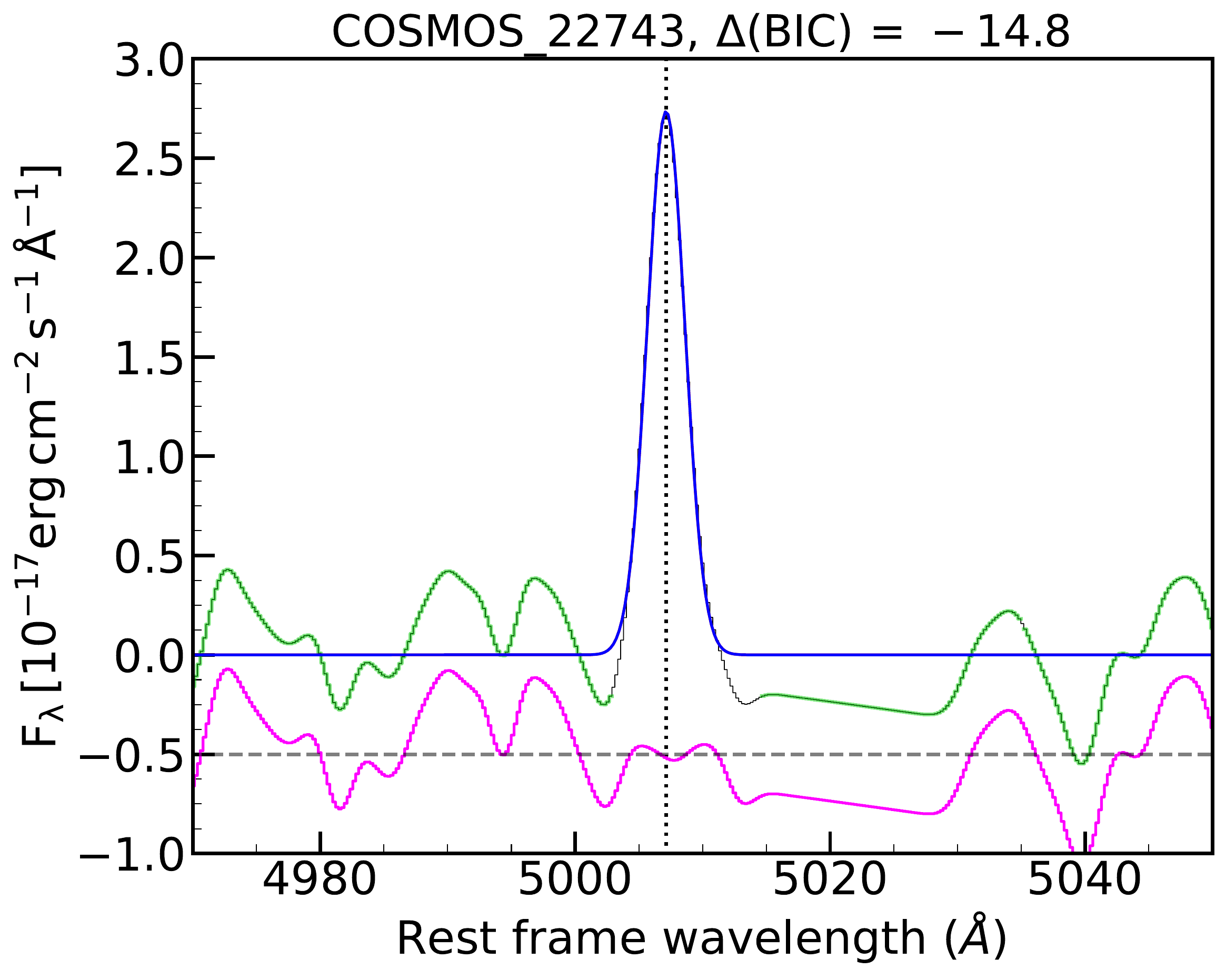}
  \caption{All spectra of our fitting. For each object, we show the spectral data (in black), the components fitted for emission lines (in blue), the overall total best-fitting model (in red), and the residuals (in magenta). The dashed vertical lines mark the location of the rest-frame  [OIII]$\rm{\lambda5007\,\AA}$ line. The first two rows are AGN, while the last two are non-AGN.}
  \label{OIIIfitting_exmple}
\end{figure*}

We show the spectral fitting for all our spectra in \autoref{OIIIfitting_exmple}, where a solid black line represents the data.\,For the spectral fitting process, we first fit the continuum, where each spectrum was visually inspected to select regions free of emission features  around the $\rm{[OIII]\lambda\,5007\,\AA}$ line; a continuum region is shown in green in \autoref{OIIIfitting_exmple}, though these regions extend beyond the depicted panel ranges.\,After this, a linear pseudo-continuum was fitted using a polynomial with a degree between four and six using all the spectral windows with those emission line free regions, and this was then subtracted from the spectrum. After the subtraction of the continuum, the [OIII]$\rm{\lambda5007\,\AA}$ line profile in the spectrum was modelled with a single or multiple Gaussian-profiles with {\sc PySpeckit}\footnote{\url{https://pyspeckit.readthedocs.io/en/latest/}}, an extensive spectroscopic analysis toolkit for astronomy, which uses a Levenberg-Marquardt minimisation algorithm \citep{Ginsburg2011}. For the Gaussian-profile fit process, we set all parameters to be free.

To prevent over-fitting a Gaussian component, and to allow for the appropriate number of Gaussians, we perform two tests. First, following \cite{Cazzoli2018}, we calculated the standard deviation of a segment of the continuum free of both emission and absorption lines ($\varepsilon_{\rm c}$). We then compared this value with the standard deviation estimated from the residuals under the [OIII]$\rm{\lambda5007\,\AA}$ emission line ($\varepsilon_{\rm line}$). We have considered a fit as reliable when $\mathrm{\varepsilon_{c}\,\approx\,\varepsilon_{line}}$, being aware that the standard deviation under the fitted line becomes smaller than the true (featureless) continuum if the emission line is over-fitted. Secondly, we use the Bayesian Information Criterion (BIC) statistic, $\mathrm{\Delta\,BIC\,=\,\left(\chi^2_{\rm 1}+k_{\rm 1} ln\left(n\right)\right) - \left(\chi^2_{\rm 2}+k_{\rm 2} ln \left(n\right)\right)}$, where $\chi^2_{\rm 1,2}$ are the total $\chi^2$ of the line fit from the single and double Gaussian, $k_{\rm 1,2}$ are the number of degrees of free parameters in the fit, and $n$ is the number of data points. Typically $\rm{\Delta}$\,BIC$\rm{\,>}$10 (see $\rm{\Delta}$\,BIC value in \autoref{OIIIfitting_exmple}) is used as strong evidence that a double component is required \citep[e.g.,][]{Swinbank2019,Avery2021,Vietri2022}.

As shown in \autoref{OIIIfitting_exmple} (blue lines), our process gives different fitted components for the [OIII]$\rm{\lambda5007\,\AA}$ emission line. The overall total best-fit model (red line) and the residuals (magenta line below) are also shown.

Generally, we consider as best fit a case when both of the two conditions mentioned above are fulfilled. In four of the AGN (COSMOS\_17759, COSMOS\_29436, COSMOS\_20881, and COSMOS\_11702), two Gaussians per line profile are required for a satisfactory fit. %(see 
Only in one AGN (COSMOS\_8450) is an additional Gaussian component required to reproduce the observed line profile (i.e., a total of three kinematic components). The remaining three AGN (COSMOS\_28917, COSMOS\_9239 and COSMOS\_30402) are well fitted with a single Gaussian component. All non-AGN are well reproduced with a single Gaussian component. Hence, we found that AGN have more complex [OIII]$\rm{\lambda5007\,\AA}$ line profiles.

\subsection{Error estimation}
In order to estimate uncertainties of all measured parameters, synthetic spectra were created from the original spectra.
We randomly generated 100 spectra for each galaxy, by adding a value (positive or negative) to each data point in the original spectrum, drawn from a Gaussian normal distribution based on the standard errors of each data point at that location in the original spectra.  Using the newly created random spectra, we repeated the entire fitting procedure with pseudo-continuum subtraction and line emission component fitting. We then estimated the uncertainty of the fitted parameters by computing the standard deviation of the relevant parameter from the fits to the 100 realisations of the random galaxy spectra. To ensure the consistency of the procedure, we in addition visually verified each emission line fitting. The estimated errors are given in \autoref{Table:FWHMparemeters},and are of order of 2\% $-$ 4\% for the wing component and 1$-$7\% for the core component.  

\section{Results}\label{sec:results}

As explained in \autoref{fittingprocess}, five out of eight spectra show a broad/blue wing, hence their modelling requires two kinematic components. We decomposed the [OIII]$\rm{\lambda5007\,\AA}$ line into a core component $\rm{([OIII]_{cc})}$ and a wing component referred to as a blue wing $\rm{([OIII]_{bc})}$. We measured the velocity using full width at half maximum (FWHM) to analyse both components. 

The FWHM of the Gaussian components are corrected for instrumental SALT FWHM following the formula: $\rm{FWHM_{cor}\,=\,(FWHM^{2} _{obs}\,-\,FWHM^{2}_{inst})^{1/2}}$,\,where $\rm{FWHM_{obs}}$ is the Gaussian FWHM measured from the spectrum, $\rm{FWHM_{inst}}$ is the instrumental FWHM, and $\rm{FWHM_{cor}}$ is the corrected value used in the analysis. 

\begin{figure*}[!ht]
\centering
\begin{minipage}[c]{.49\textwidth}
\includegraphics[width=0.89\textwidth,angle=0]{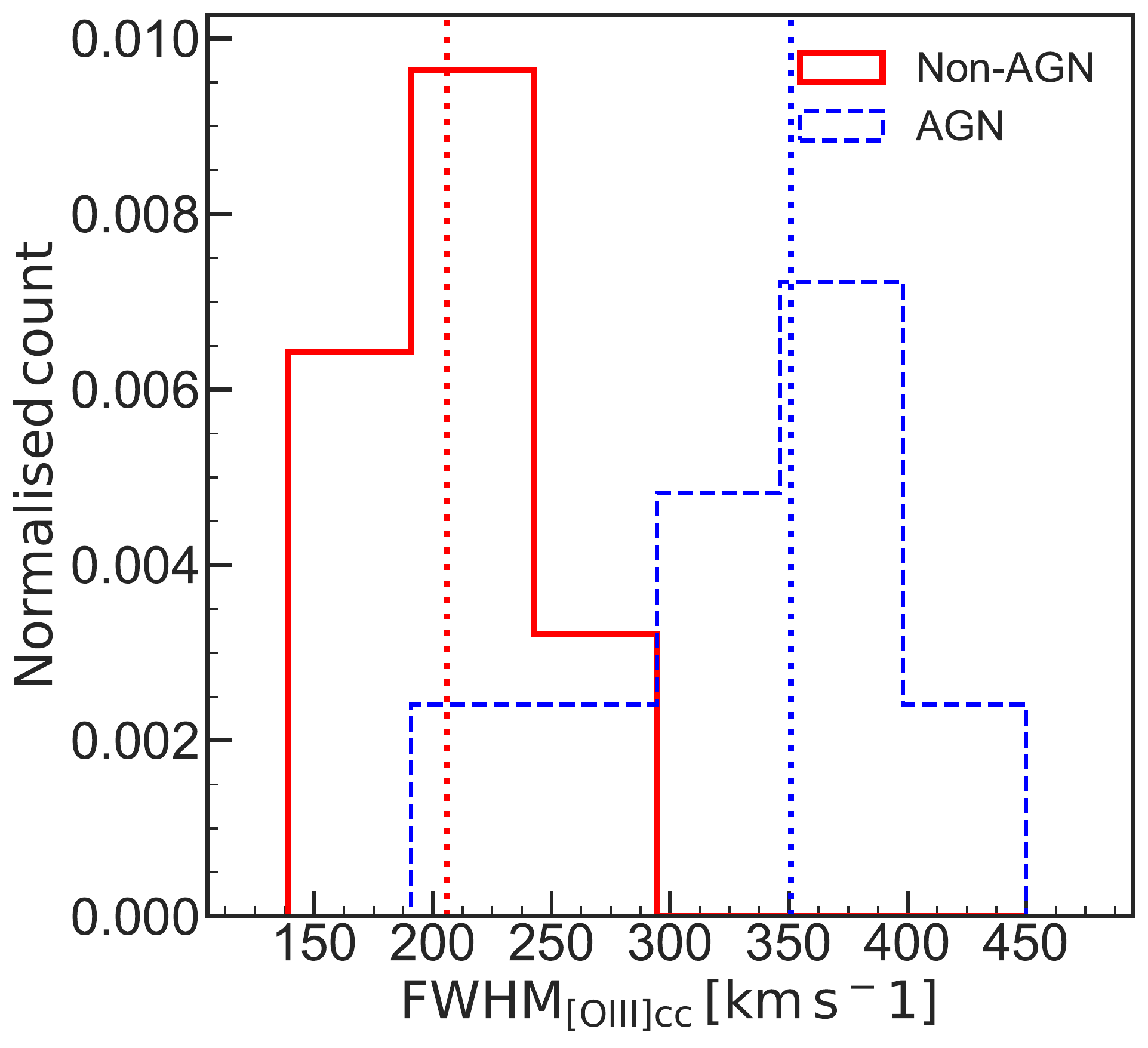}
\end{minipage}
\begin{minipage}[c]{.49\textwidth}
\includegraphics[width=0.89\textwidth,angle=0]{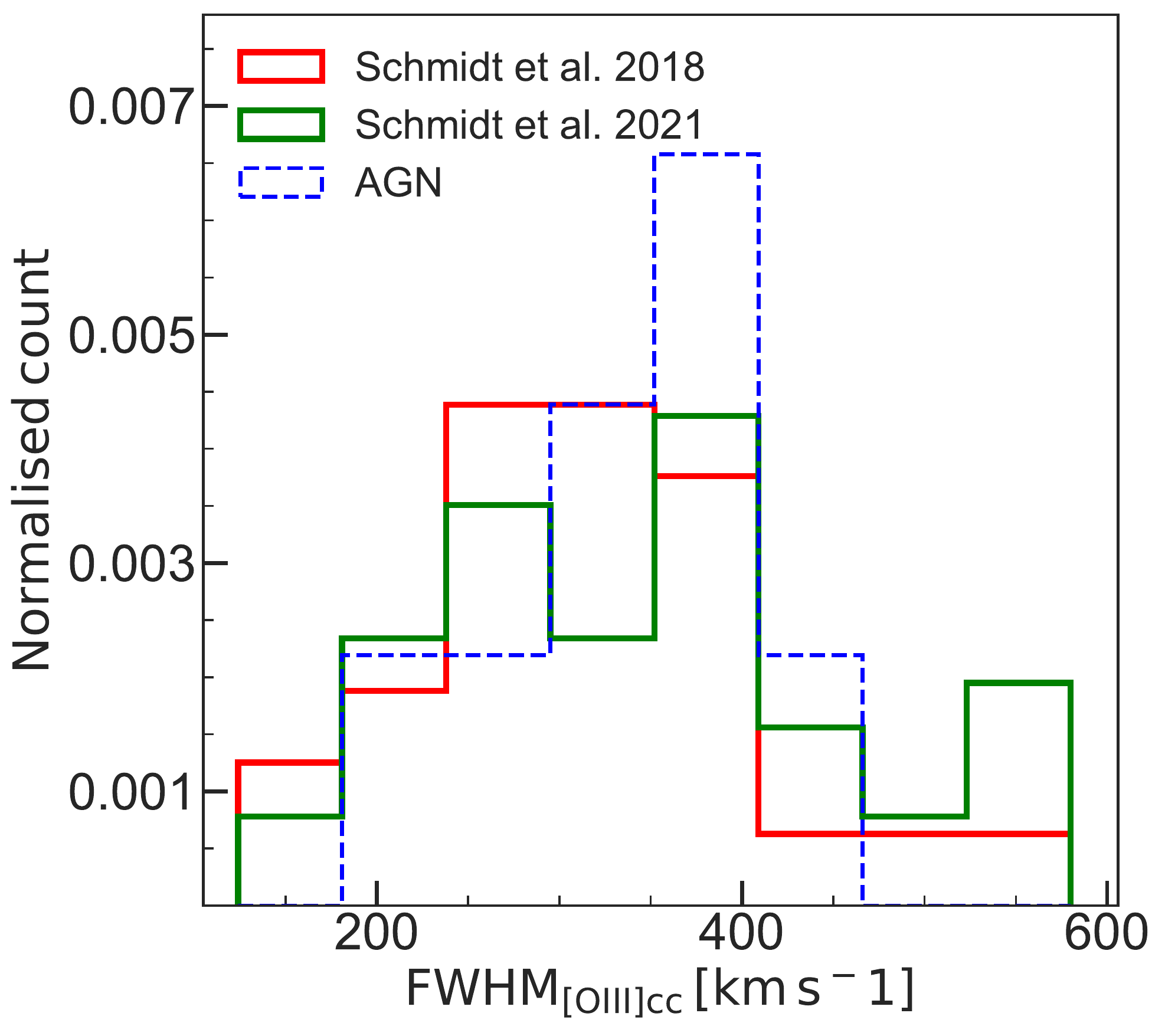}
\end{minipage}
\caption{{\em Left}: Normalised distributions of $\rm{FWHM_{[OIII]cc}}$ velocity of our AGN (dashed blue lines) and non-AGN (solid red lines). The vertical dashed lines show the median values for each histogram, using the same colour style. {\em Right}: Normalised distributions of $\rm{FWHM_{[OIII]cc}}$ velocity of our AGN (dashed blue lines), and the comparison samples of \cite{Schmidt2018, Schmidt2021} (solid green and red lines, respectively).}
\label{OIIIcc}
\end{figure*}

In \autoref{OIIIcc}, we show the FWHM distribution of measured $\rm{[OIII]_{cc}}$ and $\rm{[OIII]_{bc}}$ for AGN and non-AGN. \autoref{Table:FWHMparemeters} presents the line widths of the FWHM [OIII]$\rm{\lambda5007\,\AA}$ components and their errors (columns 2\,-\,4). All tabulated FWHM were corrected for the instrumental broadening, as explained above.
\vspace{-4mm}

\begin{deluxetable*}{lccccc}
\tablecolumns{6}
\tablecaption{[OIII]$\rm{\lambda5007\,\AA}$ emission line properties. The mid-horizontal line separates AGN (located above) and non-AGN galaxies (located below). The "bc" subscript denotes the blue-shifted component, "cc" core component and "rc" the redshifted component\label{Table:FWHMparemeters}.}
\tablehead{
\colhead{ID}                  &
\colhead{$\rm{FWHM_{bc}}$}    &
\colhead{$\rm{FWHM_{cc}}$}    &
\colhead{$\rm{FWHM_{rc}}$}    &
\colhead{$\rm{\Delta V}$}     &
\colhead{$\rm{V_{max}}$}      \\
 \colhead{}                   &
\colhead{($\rm{km\,s^{-1}}$)} &
\colhead{($\rm{km\,s^{-1}}$)} &
\colhead{($\rm{km\,s^{-1}}$)} &
\colhead{($\rm{km\,s^{-1}}$)} &
\colhead{($\rm{km\,s^{-1}}$)}    
}
\startdata
COSMOS\_17759   & 353 $\pm$15         & 264 $\pm$5         & --               & $-$169 $\pm$18      & 468 $\pm$22     \\             
  COSMOS\_29436 & 832 $\pm$34        & 383 $\pm$15      & --               & $-$94 $\pm$25       & 801 $\pm$32     \\              
  COSMOS\_20881 & 597 $\pm$17        & 334 $\pm$4        & --               & $-$262 $\pm$16       & 769 $\pm$19     \\              
  COSMOS\_8450  & 812 $\pm$31       & 450 $\pm$12        & 640 $\pm$43       & $-$351 $\pm$28       & 1041 $\pm$12    \\            
  COSMOS\_11702 & 372 $\pm$16        & 234 $\pm$11        & --               & $-$157 $\pm$10      & 474 $\pm$13      \\               
  COSMOS\_28917 & --                 & 333 $\pm$2         & --               & --                  & --              \\           
  COSMOS\_9239  & --                 & 398 $\pm$9         & --               & --                  & --              \\            
  COSMOS\_30402 & --                 & 368 $\pm$4         & --               & --                  & --              \\ 
  \hline        
  COSMOS\_6026  & --                 & 235 $\pm$13       & --               & --                  & --              \\           
  COSMOS\_7952  & --                 & 219 $\pm$3        & --               & --                  & --               \\       
  COSMOS\_19854 & --                 & 189 $\pm$2         & --               & --                  & --             \\ 
  COSMOS\_21658 & --                 & 139 $\pm$10        & --               & --                  & --              \\       
  COSMOS\_21854 & --                 & 269 $\pm$4         & --               & --                  & --              \\     
  COSMOS\_22743 & --                 & 192 $\pm$2        & --               & --                  & --                   
\enddata
\end{deluxetable*}
\vspace{-4mm}
It can be seen from \autoref{Table:FWHMparemeters} that our AGN core components $\rm{[OIII]_{cc}}$ have higher velocities than in non-AGN (see also \autoref{OIIIcc} left panel). The AGN sources cover a range of $\rm{FWHM_{cc}}$ from 234 to 450 $\rm{km\,s^{-1}}$, with a median velocity of 350 $\rm{km\,s^{-1}}$, while the respective range for the non-AGN covers velocities of 139$-$269 $\rm{km\,s^{-1}}$, with a median velocity of 205 $\rm{km\,s^{-1}}$.

In \autoref{OIIIcc} (right panel), for the AGN sample, we compared our $\rm{[OIII]_{cc}}$ FWHM  measurements (where available) with those of \cite{Schmidt2018, Schmidt2021}. The study of \cite{Schmidt2018}, with 28 narrow-line Seyfert 1 galaxies at $\rm{z\,<\,0.15}$, and of \cite{Schmidt2021}, with 45 broad-line Seyfert 1 galaxies at $\rm{z\,<\,0.06}$, is concentrated on [OIII]$\rm{\lambda 5007\,\AA}$ line profiles related to outflows. The authors confirmed the correlation between the blueshift and FWHM of the line core, between the outflow velocity and the black hole mass using a multi-Gaussian component fitting process. It can be seen that our AGN \textbf{$\rm{[OIII]_{cc}}$} measurements cover the same range as \cite{Schmidt2018, Schmidt2021}. Our measurements are consistent with also other works focused on NLR emission line studies \citep[e.g.,][]{Osterbrock1989, RodrguezArdila2000, Boroson2005, Schmidt2016}.

\begin{figure}[!ht]
\centering
\includegraphics[width=\columnwidth]{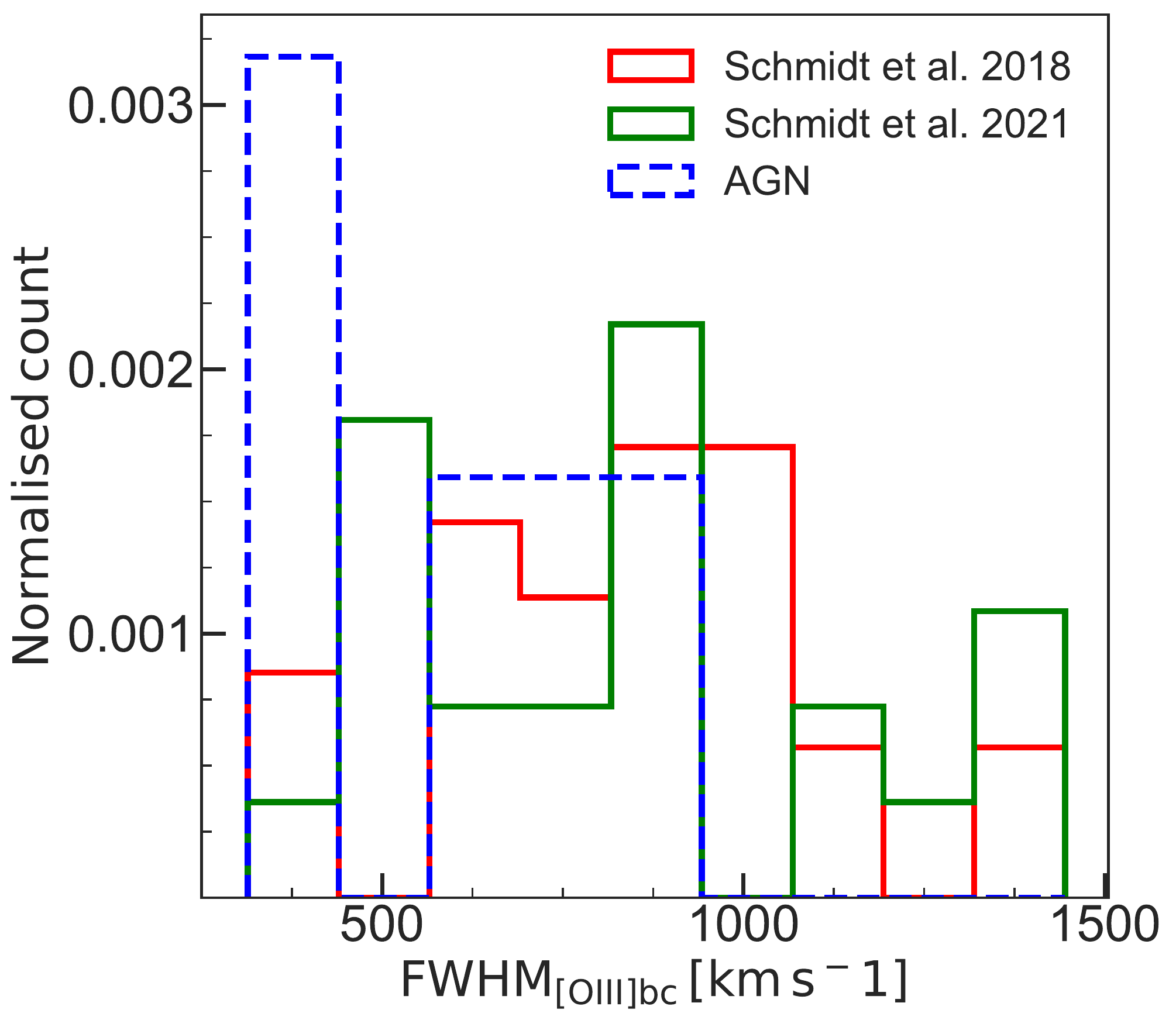}
\caption{Comparison of normalised distributions of the widths of the blue components, FWHM [OIII]bc velocities, in our sample, and in the samples of \cite{Schmidt2018,Schmidt2021}.}
\label{OIIIbc}
\end{figure}

The comparison of the widths of the blue wings, i.e. our FWHM $\rm{[OIII]_{bc}}$ velocities compared with the FWHM  $\rm{[OIII]_{bc}}$ velocities from \cite{Schmidt2018, Schmidt2021}, is shown in \autoref{OIIIbc}. We found that our AGN FWHM $\rm{[OIII]_{bc}}$ velocities range from is 353 to 832 $\rm{km\,s^{-1}}$, with a FWHM median velocity of 597 $\rm{km\,s^{-1}}$. These values are lower than those obtained by \cite{Schmidt2018} and \cite{Schmidt2021}, where the 50\% ranges of FWHM $\rm{[OIII]_{bc}}$ are $\rm{668-1007}$\, $\rm{km\,s^{-1}}$ and  $\rm 561-1069$\,$\rm km\,s^{-1}$, respectively. This could be due to the fact that our sample consists of type-2 AGN, compared to type-1 AGN analysed by \cite{Schmidt2018, Schmidt2021}.

\subsection{Kinematics from the $\rm{[OIII]\lambda\,5007\,\AA}$ emission line}
The spectral fitting of $\rm{[OIII]\lambda\,5007\,\AA}$ explained in \autoref{fittingprocess} shows that double-Gaussian decomposition (core and blue wing components) was required for five out of the eight AGN. 
To quantify the velocity shift of the wing relative to the core component in these sources, we computed the difference between core $\rm{[OIII]_{cc}}$ and wing $\rm{[OIII]_{bc}}$ line centroids, defined as $\rm{ \Delta V\,=\,\big(\big(\lambda[OIII]_{cc}-\lambda[OIII]_{bc}\big)/\lambda[OIII]_{cc}\big)\times c}$.

We found that the five AGN show blue wings in the $\rm{\Delta V}$ range from -90 to -350 $\rm{km\,s^{-1}}$ (see \autoref{Table:FWHMparemeters}, column 5). These values are well within the typical ranges reported in the literature \citep[e.g.,][]{VeronCetty2001,Cracco2016,Boroson2005,Schmidt2018, Cooke2020,Schmidt2021}, although in some cases they are narrower that what was commonly reported, which is however expected taking into account a small number of sources in our sample. Typical uncertainty in our velocity determination is approximately $\rm{10\,-\,30\,km\,s^{-1}}$.

We also examined the relationship between $\rm{\Delta V}$ and the FWHM of the $\rm{[OIII]_{cc}}$ for those galaxies showing wings. We observed a modest Pearson correlation value of $\rm{r_{p} = -0.531}$ between these two parameters. As a result, galaxies with a higher core component FWHM tend to have more strongly blueshifted wings, which is consistent with previous findings \citep[e.g.,][]{Ludwig2012,Berton2016}.

\subsection{[OIII] luminosity versus X-ray luminosity}\label{sec:Xrayoptical}

\begin{figure}[!ht]
\centering
\includegraphics[width=\columnwidth]{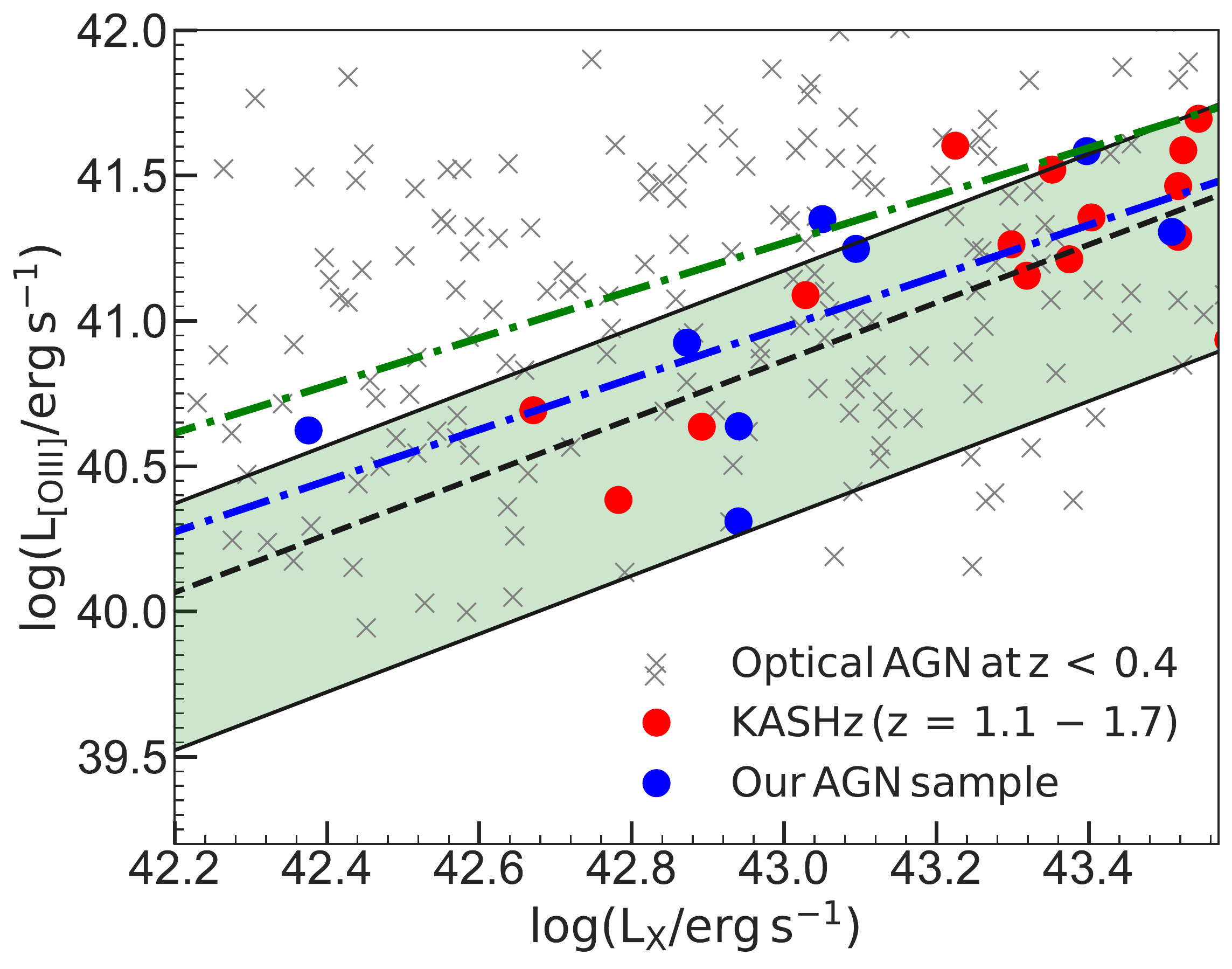}
\caption{X-ray luminosity (2--10\,keV) versus total [OIII] emission-line luminosity for our eight AGN green valley galaxies (blue dots). The blue dashed line indicates the best fit of our sample. Grey cross symbols represent the AGN at $\rm{z\,<\, 0.4}$ from \cite{Mullaney2013}. The red dots represent KASHz data, with a median value $\rm{-2.1_{-0.5}^{+0.3}}$ of  $\rm{\log(L_{{\rm [O~III]}}/L_{{\rm X}})}$, over-plotted by the black dashed line. The transparent green region between the two solid black lines indicates the $\approx$1$\sigma$ scatter on this ratio (more details are given in \cite{Harrison2016}). The green dot-dashed line shows the relationship between the local Seyferts and luminous type 1 Seyfert galaxies presented in \cite{Panessa2006}.}
\label{loiiilx}
\end{figure}

We used the [OIII] emission-line luminosity ($\rm{L_{[O III]}}$) and 2\,-\,10\,keV X-ray luminosity ($\rm{L_{X}}$) to test the $\rm{L_{X}}$ vs. $\rm{L_{[O III]}}$ correlation \citep[e.g.,][]{Caccianiga2007, Yan2011,Harrison2016, Kakkad2020}. These two quantities can be used to study the total AGN power \citep[e.g.,][]{Maiolino2003,Schmitt2003, Masegosa2011}. To measure the X-ray luminosities we used the public XMM-\textit{Newton} \citep{Brusa2007} and Chandra \citep{Civano2016, Marchesi2016} catalogues available in the COSMOS survey\footnote{https://cosmos.astro.caltech.edu/page/xray}. We extracted the X-ray fluxes in the 2\,-10\,keV band and made sure that all X-ray fluxes have reliable values with errors $<$\,20\%. X-ray luminosities were measured using the k-corrected fluxes, where the k-correction was measured as in \cite{Ptak2007} using the photon index of 1.9. \autoref{loiiilx} shows the [OIII] luminosity versus the X-ray luminosity for our AGN green valley sample. To compare our sample with previous works, we make use of the K-band Multi-Object Spectrograph (KMOS) AGN Survey at High redshift (KASHz, z $\rm{\approx\,0.6-1.7}$) taken form \cite{Harrison2016}, where a sample of $\sim$\,40 X-ray selected AGN were analysed, and from which a $\sim$\,50\% fraction with outflows was reported. For a low-redshift AGN comparison sample, we make use of \cite{Mullaney2013}, this sample contains optically selected AGN at z $<$ 0.4  from the SDSS spectroscopic database. This sample is similar to our AGN in terms of redshift (see \autoref{tab:obslog}, column 3).

In general, although we have a small number of sources, they are closely following the trend in relation between those of both low- and high-redshift AGN (see \autoref{loiiilx}). We find a median luminosity ratio of $\rm{\log(L_{[O III]}/L_{X})\, =\, -1.86}$ for our eight AGN, which is fully consistent with the value obtained at lower redshift. 

\subsection{Total asymmetry}
The shift of the whole $\rm{[OIII]\lambda\,5007\,\AA}$ emission line, or the "blue emission", which quantifies the full blue extension of the wing component in all our AGN with $\rm{[OIII]\lambda\,5007\,\AA}$ line profile with double-Gaussian decomposition, is one of the properties in which we are most interested. To measure the blue emission, we employed the \cite{Schmidt2018, Schmidt2021} definition:
\begin{equation}
\rm{blue\,emission\,=\,\Delta V\,-FWHM[OIII]_{bc}}
\end{equation}
The obtained blue emission values for our sample range from -1117 to -521 $\rm{km\,s^{-1}}$, with a mean value of -790 $\rm{km\,s^{-1}}$. This range is similar to that in \cite{Schmidt2018, Schmidt2021}, who find a range from -1674 to -469 $\rm{km\,s^{-1}}$.
\vskip 2.2cm

\subsection{Ionised outflow detection}

Previous works suggested that ionised winds have a non-Gaussian profile in the $\rm{[OIII]\lambda\,5007\,\AA}$ emission line, with somewhat greater wings/tails than a Gaussian \citep[e.g.,][]{Zakamska2016, Xu2020,Li2022}. The profiles often include a blue (or in rare cases, red) wing, indicating that the line profile is recreated with two components, a narrow one linked to the narrow-line region (NLR) emission, and the other wide component, that may be shifted. This second component is regarded as an outflow candidate in this work.

Furthermore, we adopted the \cite{Rupke2013} and \cite{Bischetti2017} methodology to determine a maximum outflow velocity, considering that the outflow expands at a constant velocity:
\begin{equation}
\rm{V_{max}\,=\, |\Delta V|\,+\,2\sigma_{[OIII]_{bc}}}
\end{equation}
where $\rm{\sigma_{[OIII]_{bc}}}$ is the velocity dispersion of the Gaussian representing the wing component of the $\rm{[OIII]\lambda\,5007\,\AA}$ emission line. 

Using these definitions, we calculated the maximum outflow velocities $\rm{V_{max}}$ for those five AGN galaxies exhibiting wings, out of the eight total observed, to span the range of $\rm{470\,-\,1040\,kms^{-1}}$ (see \autoref{Table:FWHMparemeters}, column 6).  

\section{Discussion}\label{sect:discussion}
In this study we present detailed observations of the $\rm{[OIII]\lambda\,5007\,\AA}$ emission line for a small sample of AGN and non-AGN galaxies with FIR emission in the green valley at $z<$ 0.4, and argue that their properties are consistent with previous works \citep[e.g.,][]{Mullaney2013, Harrison2016, Schmidt2018, Schmidt2021}.
We demonstrate that the $\rm{[OIII]\lambda\,5007\,\AA}$ line profile of the green valley FIR AGN are different from those of equivalent FIR non-AGN galaxies, with respect to the complexity of their line profiles (see \autoref{fittingprocess}), as well as the 
core component velocity width $\rm{FWHM_{[OIII]cc}}$ (see \autoref{sec:results}).  Our results suggest a possible presence of gas outflows in the AGN sample.

In order to study any characteristics of the outflowing gas, it is however important to distinguish between gravitational and outflow processes. To do this, we first compare to a conservative velocity threshold of $\rm{V_{max}\,=\,650\,km\,s^{-1}}$, as used in other works such as in \cite{Perna2017} and \cite{Rojas2020}. The value was chosen, therein, following the fact that 95\% of BOSS galaxies at $\rm{z\,<\,0.8}$ had characteristics matching this limit. 
 Using this $\rm{V_{max}}$ criterion, only three of our AGN would be classified as "true" outflow signals (COSMOS\_29436, COSMOS\_20881, and COSMOS\_8450).  However, we note that no matching of masses of the respective samples, i.e.\ ours vs.\ the BOSS galaxies referred to, was done.  We also checked against an outflow criterion of \cite{Perna2015, Perna2015b}, where $\rm{FHWM_{bc}}$ threshold of 550 $\rm{kms^{-1}}$ was used.  The same three galaxies satisfy this limit, with $\rm{FHWM_{bc}}$ of 653 $\rm{kms^{-1}}$, 613 $\rm{kms^{-1}}$ and 790 $\rm{kms^{-1}}$, for COSMOS\_29436, COSMOS\_20881, and COSMOS\_8450, respectively.

We then use the spatial information available in our SALT spectra to gain more information regarding the origin of the motion of gas, whether nuclear, or disk rotation, or outflow.
The broadening of the observed $\rm{[OIII]\lambda\,5007\,\AA}$ line can be caused by the rotation of the galaxy, if any significantly rotating parts of it enter the slit. Therefore, it is important to extract the 1D spectra centred on the galaxy nucleus and make the observations along the minor axes, as was indeed performed.  Moreover, we compared extracted 1D spectra of various size apertures to check if the blue wings origin is spatially dependent. In addition to extracting nucleus-centered aperture 1D spectra, we also extracted apertures separately from the two sides of the galaxy nucleus.  All this was done with all AGN showing wing components (namely COSMOS\_17759, COSMOS\_29436, COSMOS\_20881, COSMOS\_8450, COSMOS\_11702). The nuclear apertures are shown in \autoref{Spailly_plot}, where we illustrate the different profiles extracted from the centres (blue boxes) and off-nuclear (magenta boxes) of galaxies.

The results of these tests were as below.  In three of the cases (COSMOS\_29436, COSMOS\_20881, and COSMOS\_11702) the blue wing in the line profile persists in all tested apertures, regardless of the size and position with respect to the nucleus. In COSMOS\_11702 the smaller apertures in fact showed a more pronounced blue wing than that seen in the larger apertures shown of \autoref{OIIIfitting_exmple}. That the blue wing is present everywhere tells us that there is outflowing material from the galaxy projected size over the whole face of it (several kpc), i.e.\ most likely originating from the nucleus towards the observer in a wide cone-like structure typical of galactic super-winds \citep[see e.g.][]{Heckman1990}.  In particular, the observed shape of the line profile when connected with the geometry, cannot be due to e.g. disk rotation. As it can be seen from the spatial information (\autoref{Spailly_plot}), we are findings that outflow extension along the minor axis are extend up to $\rm{\sim}$ 6 kpc for COSMOS\_17759, $\rm{\sim}$ 6.6 kpc for COSMOS\_29436, $\rm{\sim}$ 7 kpc for COSMOS\_20881, $\rm{\sim}$ 6.5 kpc for COSMOS\_8450, and $\rm{\sim}$ 5 kpc for COSMOS\_11702.

In the case of COSMOS\_8450, on the other hand, the closer inspection of profiles at different spatial locations showed the following: a \emph{broad} but fairly symmetric component was detected at all locations, whether small or larger aperture, and also on both sides of the nucleus.  The best fit model to these profiles sometimes comes out with three components, as shown in \autoref{OIIIfitting_exmple}, or with two components, a narrow and broad component one but without any significant relative velocity offset.  This could well be due to high velocity gas in all apertures, including inflows and outflows, but it could also be due to disk rotation signal entering the \emph{width} of the 1.25\arcsec slit (approximately 7kpc at this redshift). The implied rotational velocity is high, more than 350 $\rm{kms^{-1}}$.  However, interestingly, this galaxy is the most massive of our sample: see \autoref{SALT_main}, where this galaxy is located at approximately $10^{11.1}$ $M_{\odot}$.  The same behaviour of a broader component persisting together with a narrow one was seen in COSMOS\_17759, though the large aperture as seen in \autoref{OIIIfitting_exmple} did give us a blue wing detection (albeit the weakest and narrowest of the five). And interestingly, this galaxy is the second most massive of our AGNs, the one seen just below $10^{11.0}$ $M_{\odot}$ in \autoref{SALT_main}.  Hence, while we see signal of high velocity gas in these two galaxies, and possible outflows, we cannot rule out the possibility of rotational/gravitational effects being the reason for that signal without higher spatial resolution data.

We do also wish to highlight that in examining the morphological types of our galaxies, we see an ongoing interaction of galaxies in (at least) the cases of COSMOS\_29436 and COSMOS\_20881 (see \autoref{slit_images} and \autoref{Spailly_plot}). This could potentially affect the line profile due to tidal gravitational effects. However, the fact that the blue wings were clearly seen in the small nuclear apertures just as they were in the larger apertures, as well as on both sides of the galaxy, strongly suggests that the outflow is driven from the nucleus, and is not, in these cases, an artefact of e.g.\ tidal stream of the neigbouring galaxy entering the slit aperture. Nevertheless, more data with higher angular resolution would be beneficial to fully understand the $\rm{[OIII]\lambda\,5007\,\AA}$ line profiles of these  sources, and, furthermore, to study the intriguing role of the interaction in feeding and/or triggering the AGN activity.

Finally, we note that the three AGN, COSMOS\_29436, COSMOS\_20881, and COSMOS\_11702, that we consider the most unambiguous cases for gas outflow signatures, and indeed nuclear-driven gas flow signatures as discussed above, are also ones within or very close to the MS of star forming galaxies. They are depicted with the thick open blue circles in \autoref{SALT_main}.\\
This particular result in this paper, based on the study of the $\rm{[OIII]\lambda\,5007\,\AA}$ emission line profiles of a small sample of optically-selected green valley AGN with FIR emission, is thus very much in line with that obtained using photometric techniques of statistical samples in \cite{Mahoro2017, Mahoro2019, Mahoro2022}, where we suggested the possibility of on-going positive AGN feedback in FIR green valley galaxies. It will be interesting to further distinguish whether we indeed are witnessing positive feedback by the AGN on star formation activity of the host galaxies, or, for example, whether any SF quenching has not \emph{yet} started in these galaxies, and if so, why, and what the exact physical role of gas outflows is. To do this, we require larger AGN samples in the green valley observed with high spatial resolution. 

\section{Summary}
In order to study the $\rm{[OIII]\lambda\,5007\,\AA}$ emission line profile, we collected optical spectroscopic data for 14 sample of FIR AGN and non-AGN optically-selected green valley galaxies using 11\,m SALT telescope.  The $\rm{[OIII]\lambda\,5007\,\AA}$ emission line profile is used to investigate the presence of gas outflows.  A multi-component Gaussian fitting approach was adopted to account for the wings of the $\rm{[OIII]\lambda\,5007\,\AA}$ emission line. The outflow velocities were estimated using different established method and criteria, and in particular by studying the line profiles combined with the spatial information along the slits provided by the observations.  

In what follows, we summarise the main results of this work and highlight future prospects:
 
\begin{itemize}

\item Overall, we see that the optically-selected FIR AGN green valley galaxies display a complex $\rm{[OIII]\lambda\,5007\,\AA}$ emission line profile (at least two components are required to fit the $\rm{[OIII]\lambda\,5007\,\AA}$ line).  The profile is more complex than 
that observed in non-AGN, where the  $\rm{[OIII]\lambda\,5007\,\AA}$ emission line is well-fitted by a single component.  We thus confirm a difference in the $\rm{[OIII]\lambda\,5007\,\AA}$ line profiles between FIR AGN and non-AGN green valley galaxies. 

\item We see indications of outflows in five out of eight observed FIR AGN optically-selected green valley galaxies as determined by blue wings to their line profiles. Among these five AGN, two and three, respectively, also satisfy certain literature outflow criteria, such as $\rm{V_{max}\,>\, 650\,kms^{-1}}$ (\cite{Perna2017, Rojas2020}) and $\rm{FWHM[OIII]_{bc}\,>\,550\,kms^{-1}}$ (\cite{Perna2015, Perna2015b}).  

\item Moreover, using spatial and geometrical arguments, as studied from multiple spectral apertures in our own data, three of those five blue-wing galaxies unambiguously show centrally driven galaxy-wide gas outflows, while in the two other cases we cannot rule out with present data a contribution to a potential outflow signature from rotation (or velocity dispersion) of these two relatively massive galaxies.

\item Finally, we note that the three unambiguous cases of outflows in the FIR AGN green valley galaxies, are situated within, or just at the border, of the MS of SF galaxies.  This suggests the possibility that we are witnessing the influence of AGN on their host galaxies in this sample (such as possible positive feedback), in line with results we have reported previously in \cite{Mahoro2017, Mahoro2019, Mahoro2022}.  Nevertheless, we consider this a preliminary result, and plan further observations of larger samples, and with data using a new generation of telescopes, including other wavelengths and with higher spatial resolution (e.g., JWST/NIRSpec or ALMA).

\end{itemize}

\section*{acknowledgments}
We thank the anonymous referee for accepting to review this paper, and for giving us constructive and useful comments that improved this paper. This work was supported by the National Research Foundation of South Africa (Grant Numbers 110816 and 132016). This paper makes use of observations taken with SALT under programs 2017-2-MLT-003, 2020-1-DDT-001 and 2020-2-SCI-021. AM gratefully acknowledges financial support from the Swedish International Development Cooperation Agency (SIDA) through the International Science Programme (ISP) - Uppsala University to the University of Rwanda through the Rwanda Astrophysics, Space and Climate Science Research Group (RASCSRG). PV acknowledges support from the National Research Foundation of South Africa. MP acknowledges the support from the Space Science and Geospatial Institute under the Ethiopian Ministry of Innovation and Technology (MInT), the Spanish Ministerio de Ciencia e Innovaci\'on -
Agencia Estatal de Investigaci\'on through project PID2019-106027GB-C41, and the State Agency for Research of the Spanish MCIU through the Center of Excellence Severo Ochoa award to the Instituto de
Astrof\'isica de Andaluc\'ia (SEV-2017-0709). We are also grateful to the python, Virtual Observatory, and TOPCAT teams for making their packages freely available to the scientific community.

\bibliographystyle{aasjournal}
\bibliography{Outflows}
\clearpage
\appendix
\section{Slit positon}\label{slit_images}
Hubble Space Telescope (HST) images taken with the Advanced Camera for Surveys (ACS) in F814W band of our 14 sample galaxies. All images were generated using COSMOS cutouts\footnote{\url{https://irsa.ipac.caltech.edu/data/COSMOS/index_cutouts.html}}. Each galaxy is scaled to 15\arcsec cutout size and the placement of the 1.25\arcsec wide slit is shown in blue. 
\begin{figure*}[b]
 \centering
\includegraphics[width=.24\textwidth]{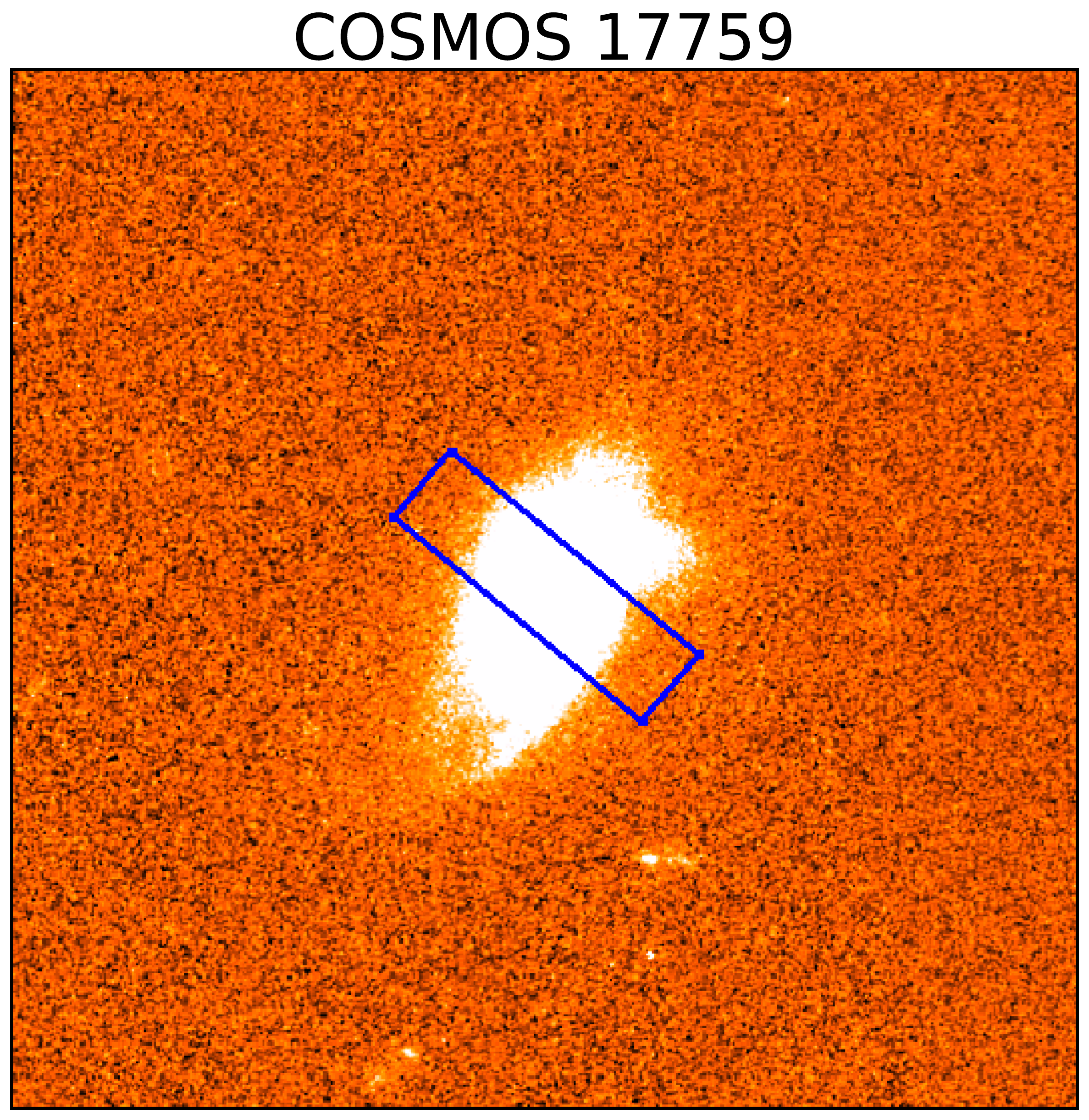}
\includegraphics[width=.24\textwidth]{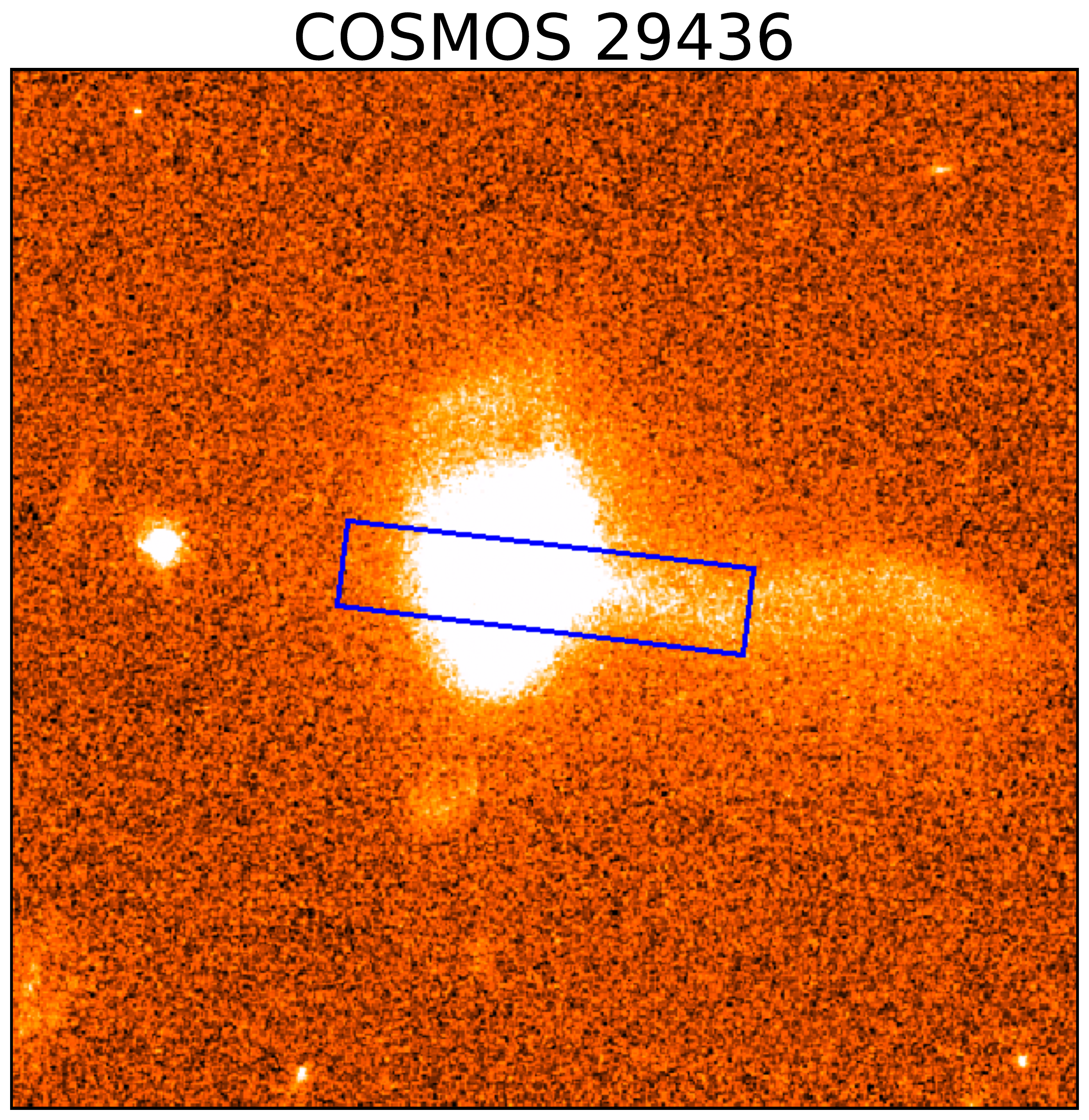}
\includegraphics[width=.24\textwidth]{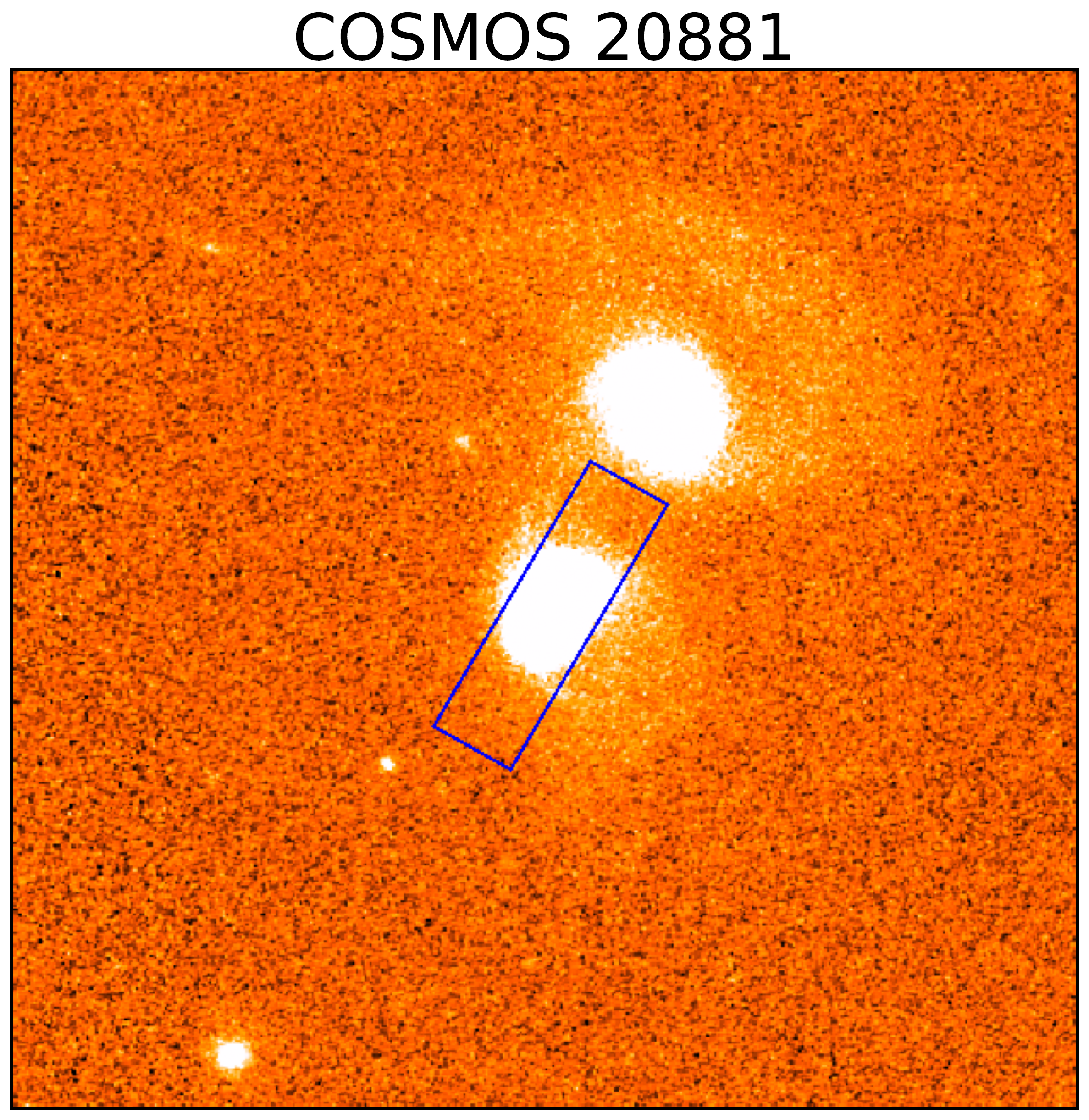} 
\includegraphics[width=.24\textwidth]{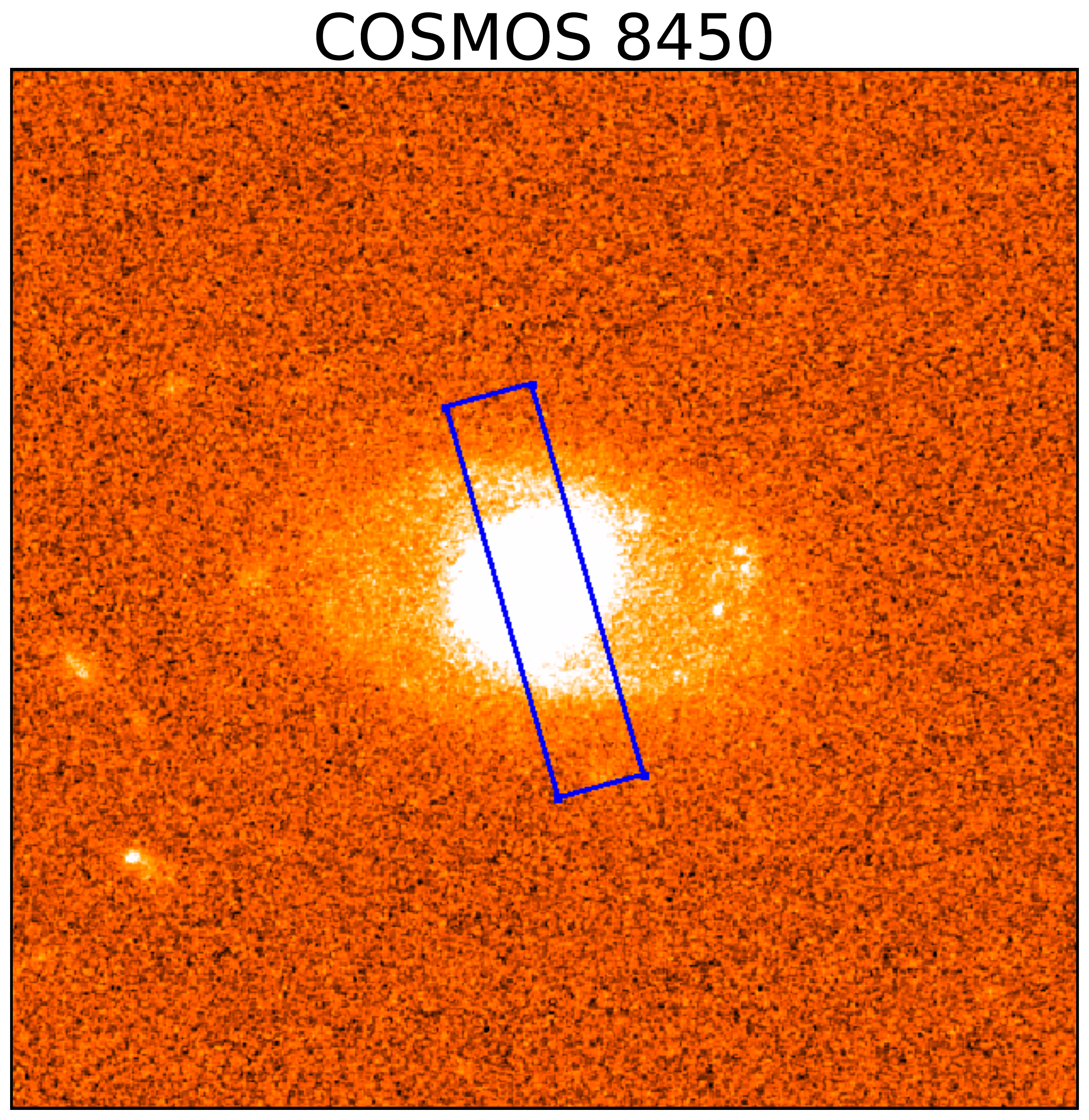}\\
\includegraphics[width=.24\textwidth]{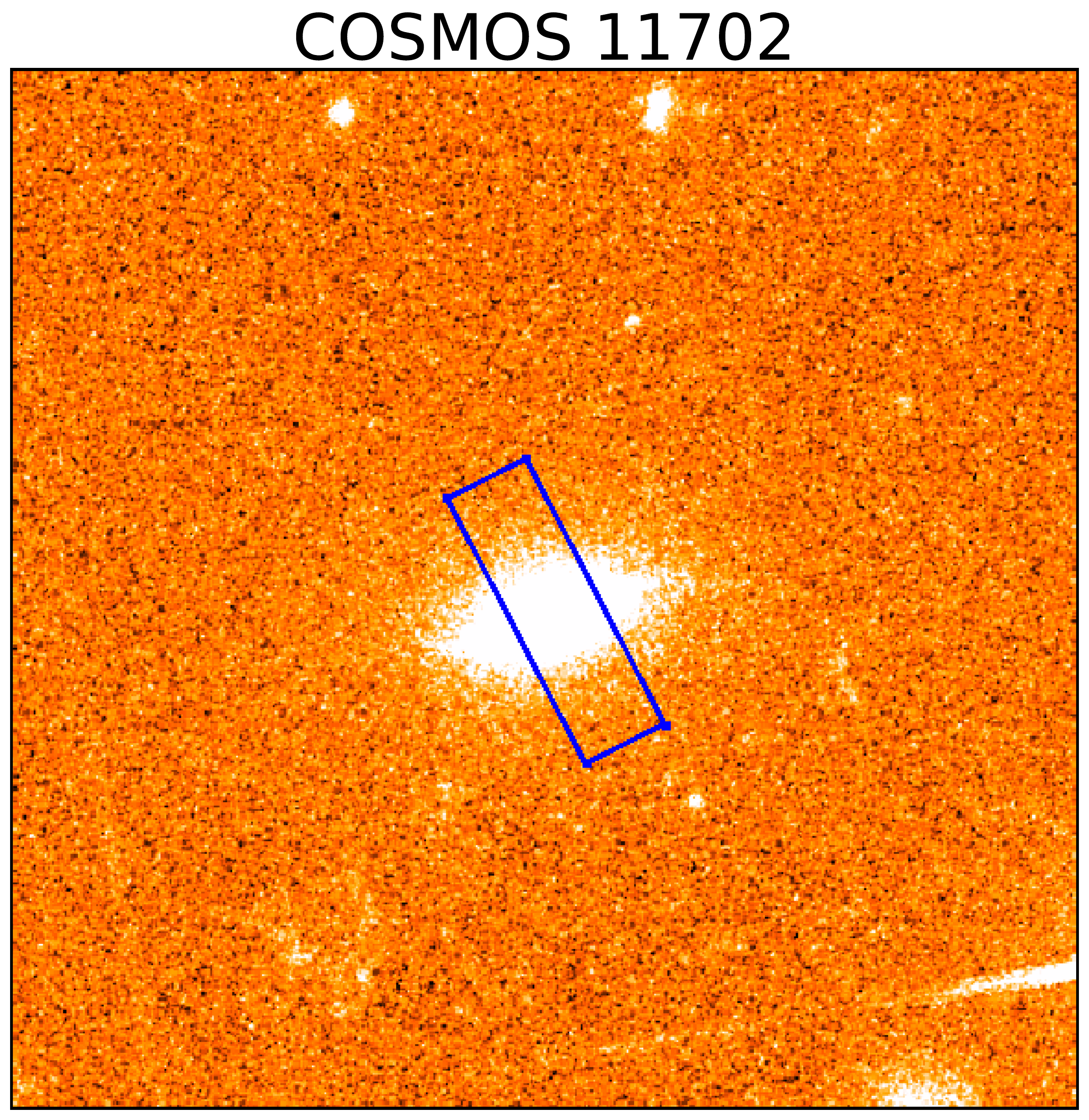}  
\includegraphics[width=.24\textwidth]{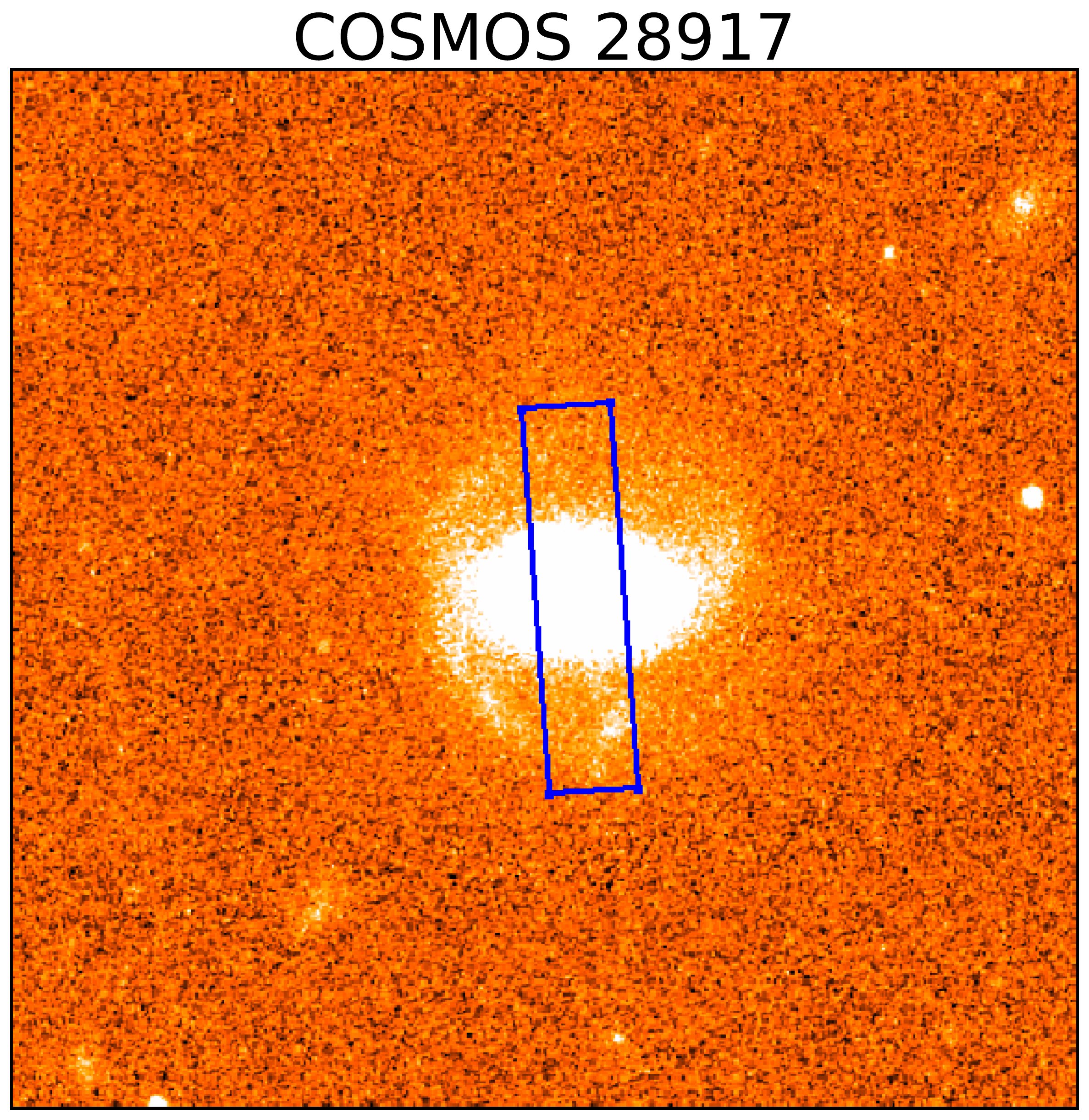}
\includegraphics[width=.24\textwidth]{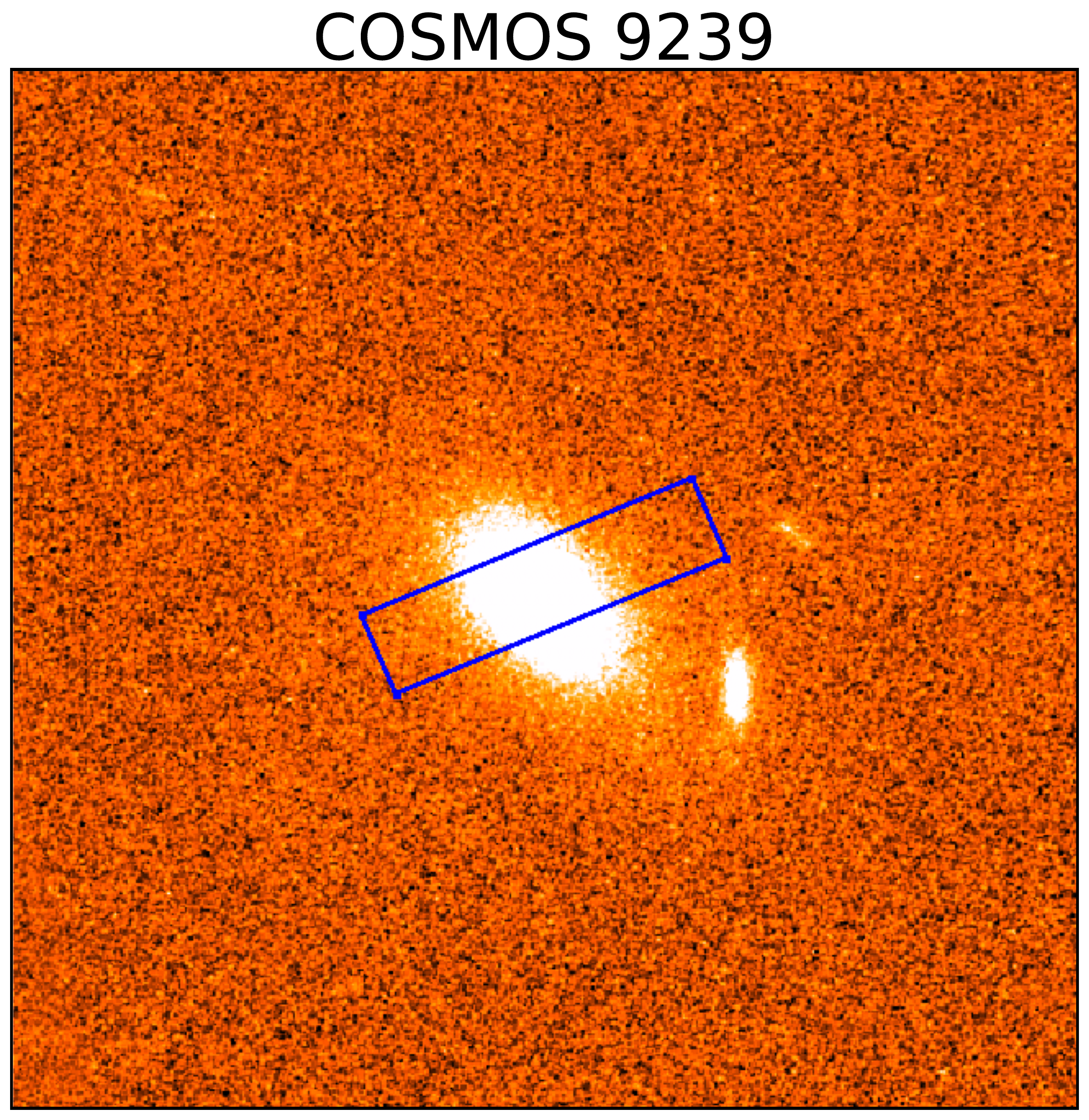} 
\includegraphics[width=.24\textwidth]{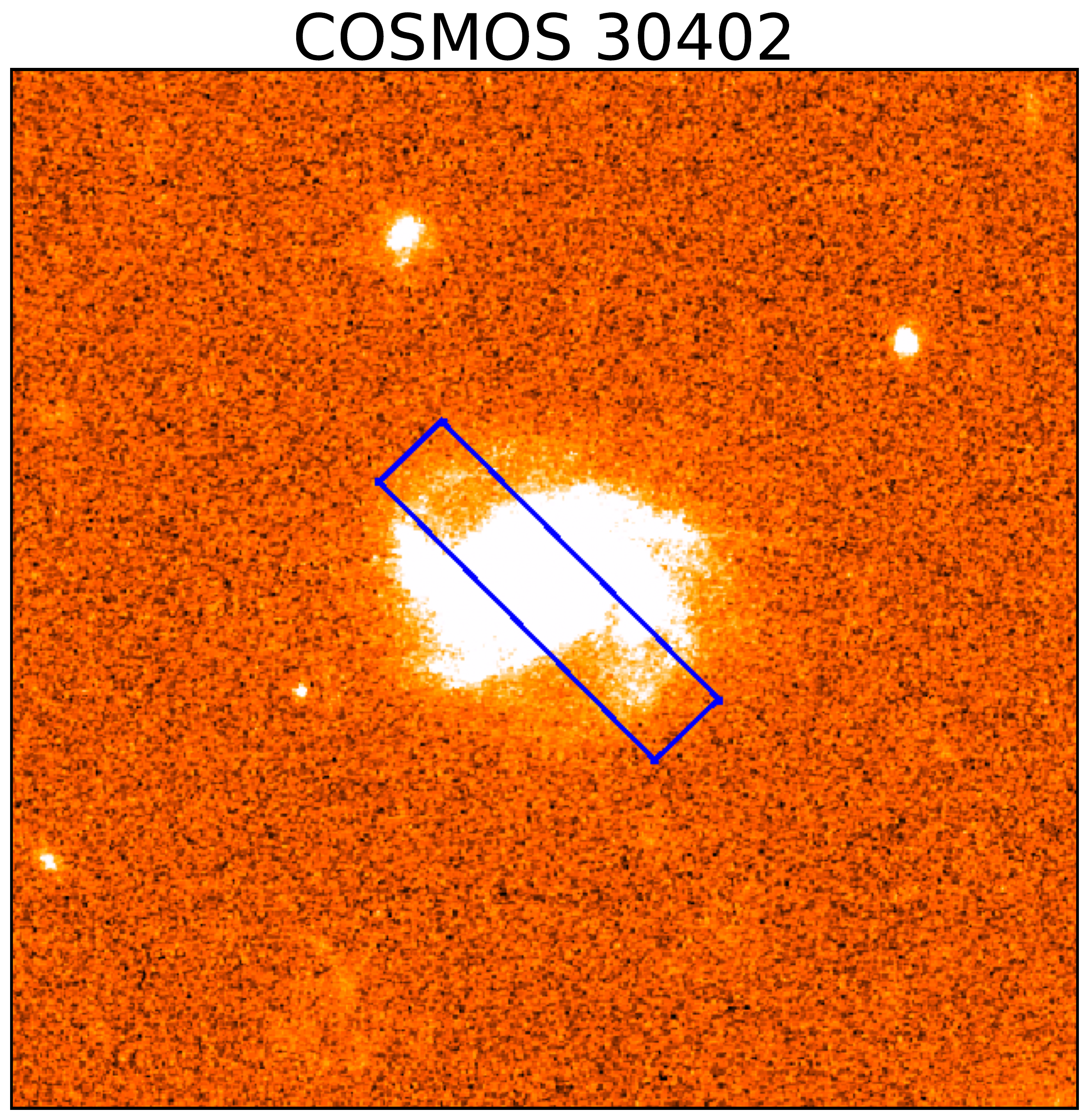}\\  \includegraphics[width=.24\textwidth]{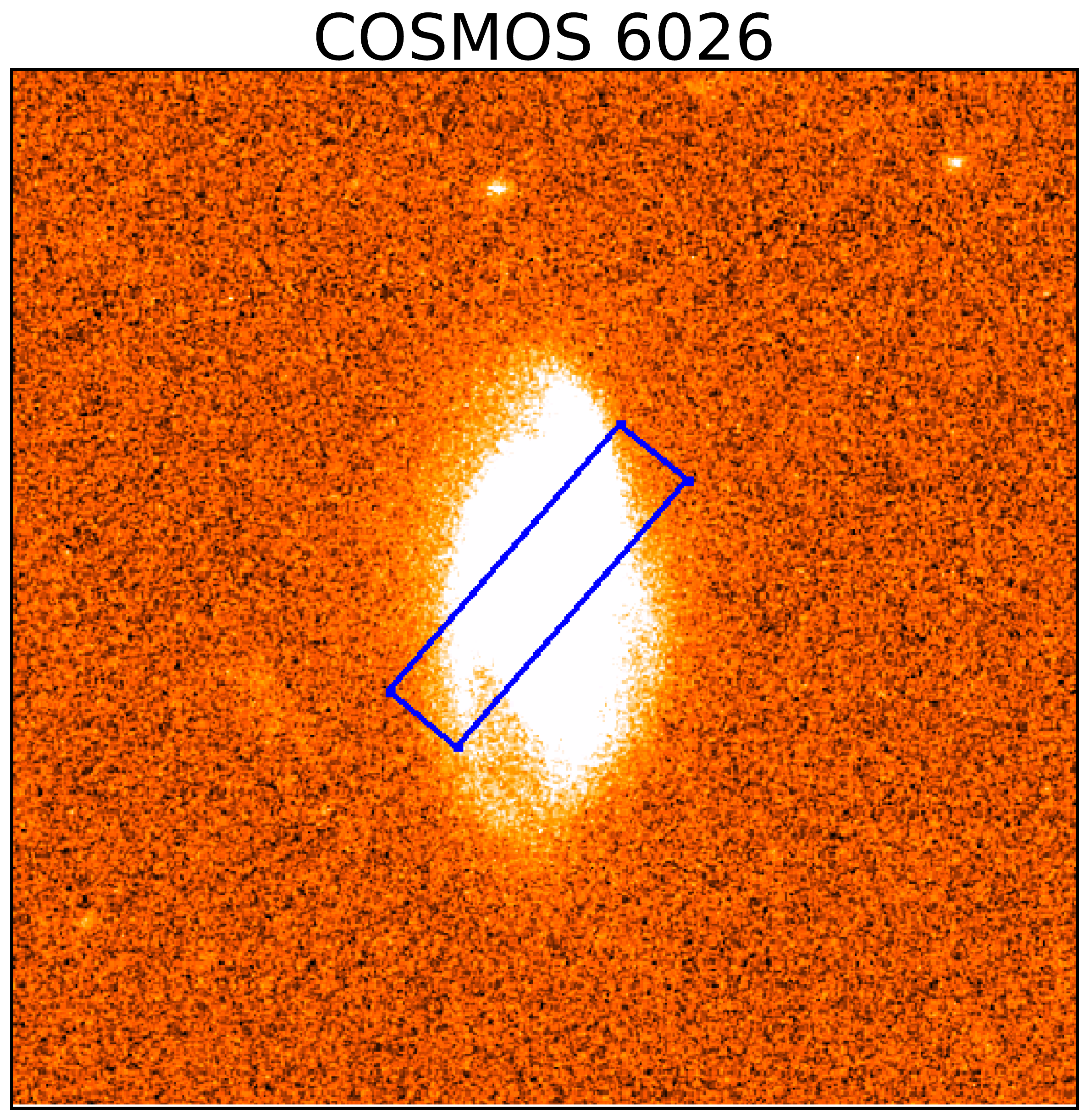}   
\includegraphics[width=.24\textwidth]{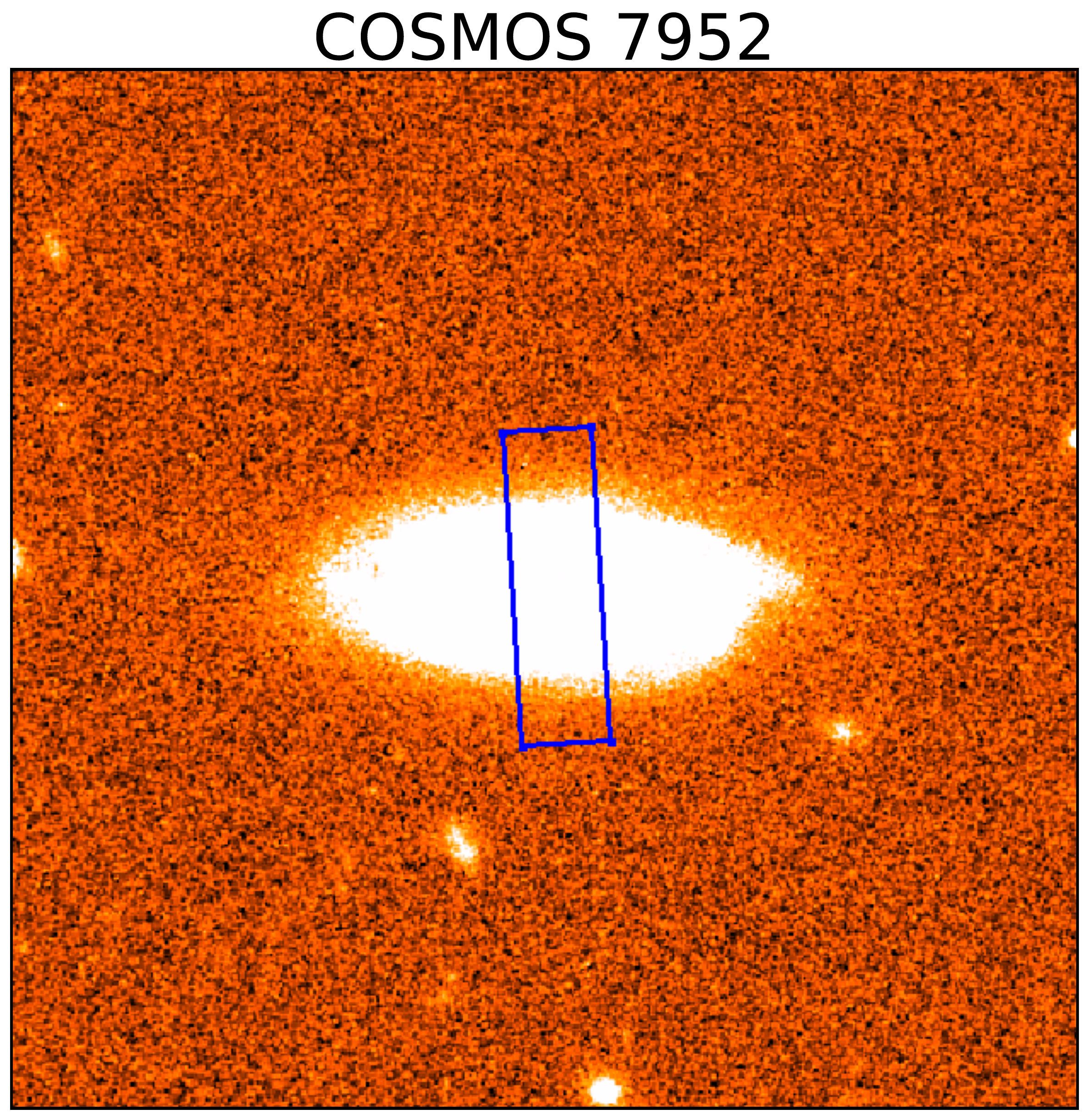}
\includegraphics[width=.24\textwidth]{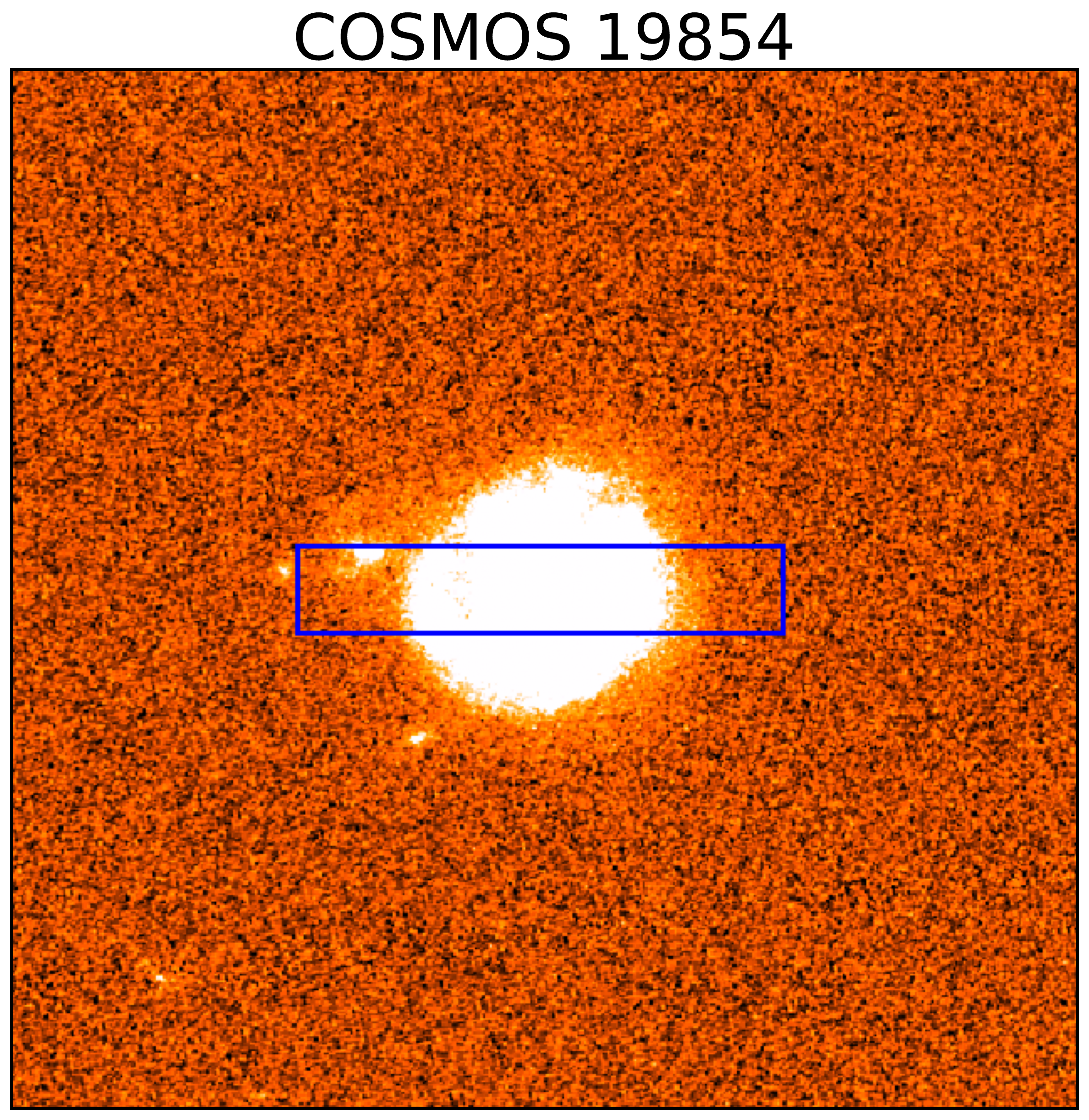}  \includegraphics[width=.24\textwidth]{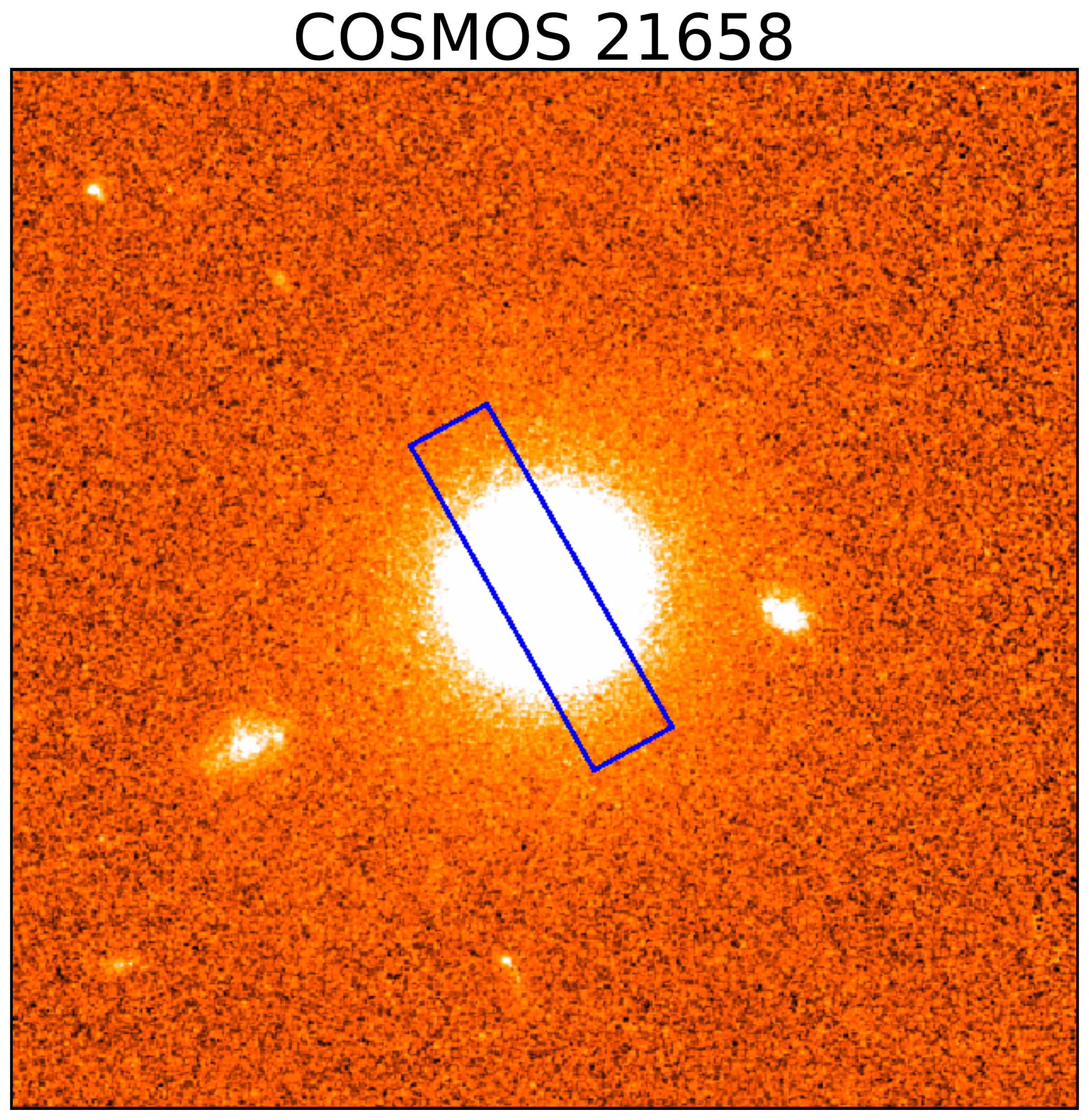} \\  
\includegraphics[width=.24\textwidth]{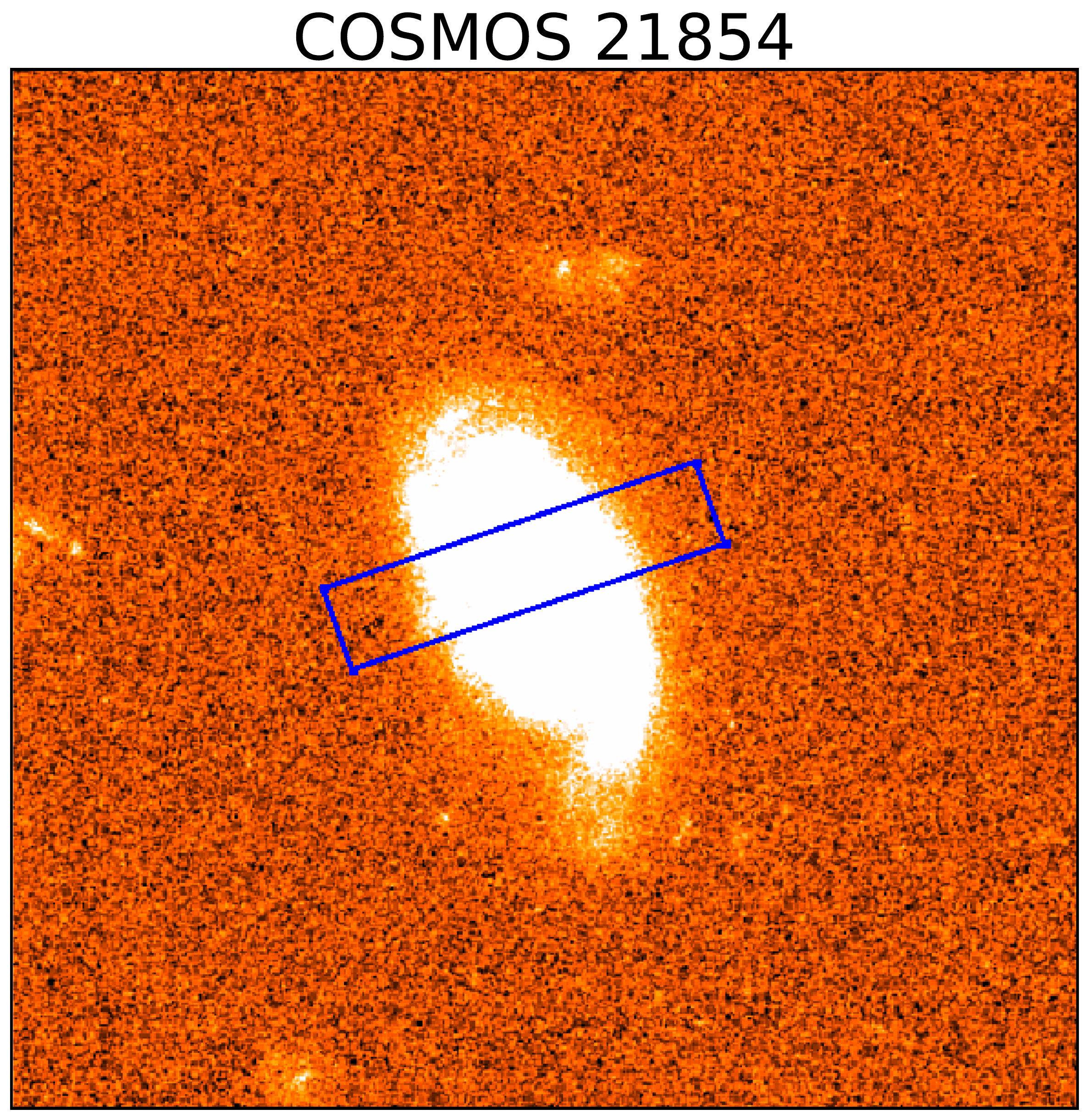} 
\includegraphics[width=.24\textwidth]{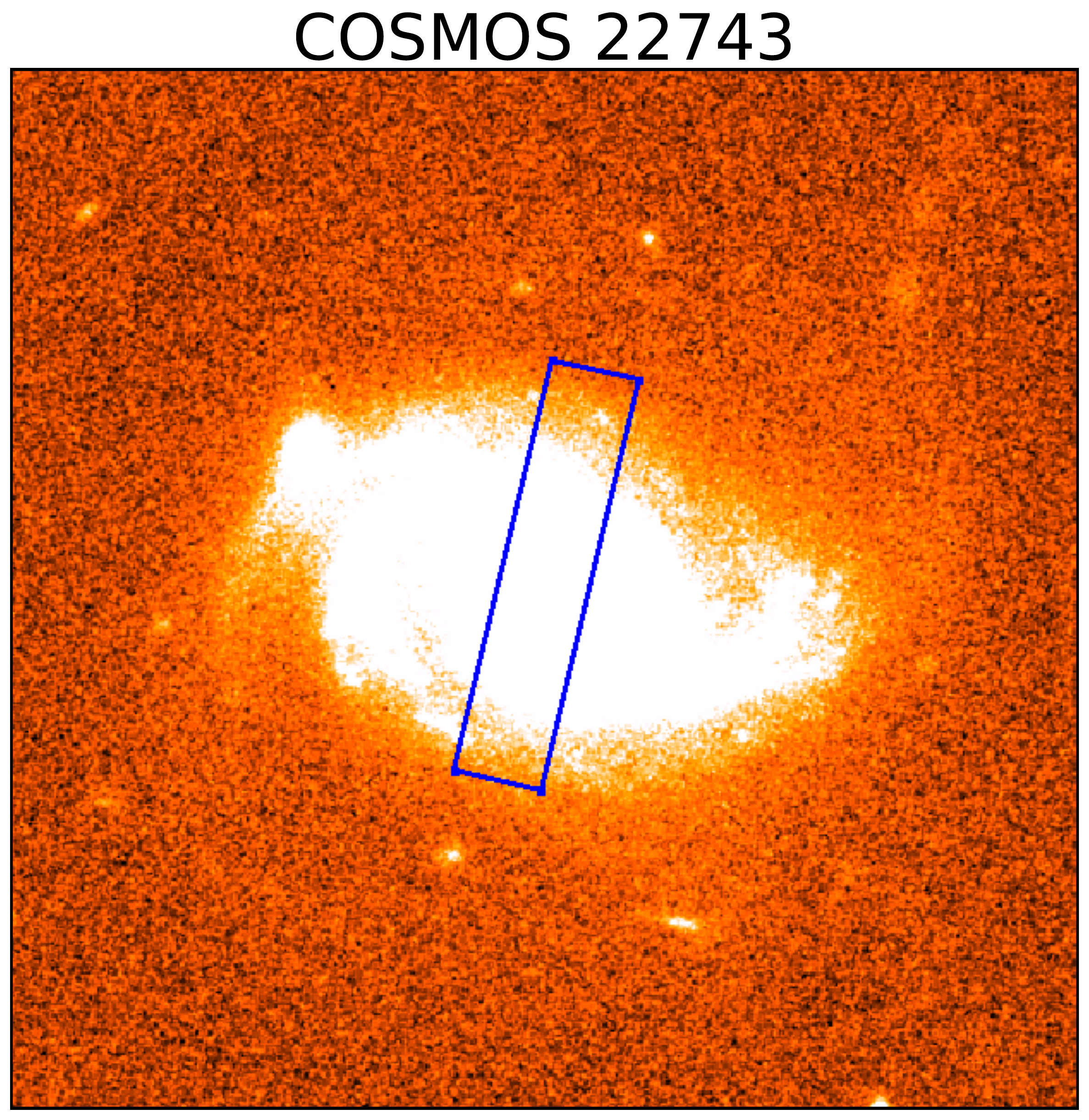}

\end{figure*}
\section{Illustration of differences in AGN line profiles extracted from the centre of a galaxy}\label{Spailly_plot}
A comparison between line profiles extracted at $\rm{\sim}$3 kpc to $\rm{\sim}$7 kpc from the center COSMOS\_17759, COSMOS\_29436, COSMOS\_20881, COSMOS\_8450 and  COSMOS\_11702 is shown. The blue boxes at the centres of AGN images show the regions where nuclear-only apertures were extracted and the magenta boxes represent the off-nuclear extraction regions. 

\begin{figure*}[h]
 \centering
\includegraphics[angle=270, width=.49\textwidth]{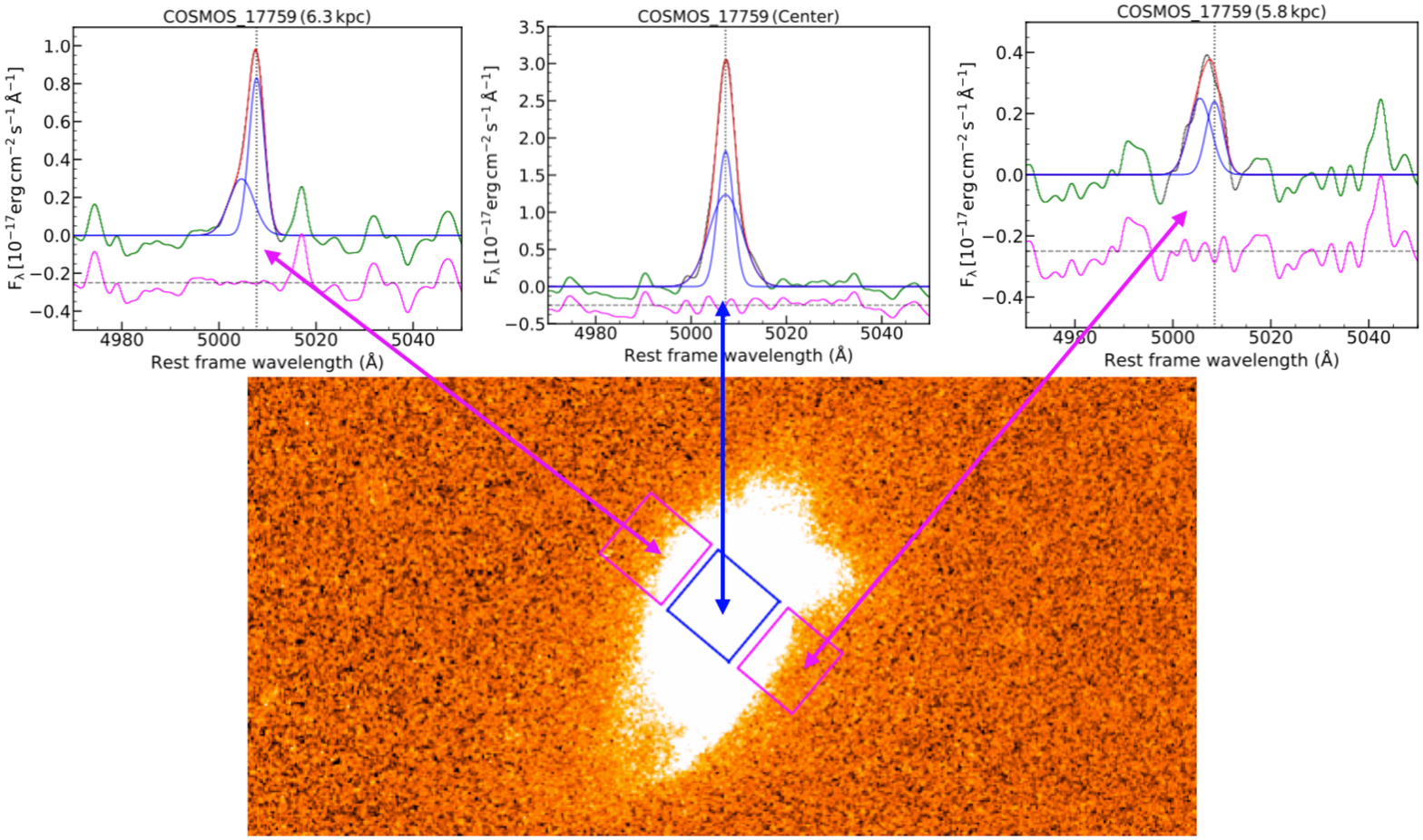}
\includegraphics[angle=270, width=.49\textwidth]{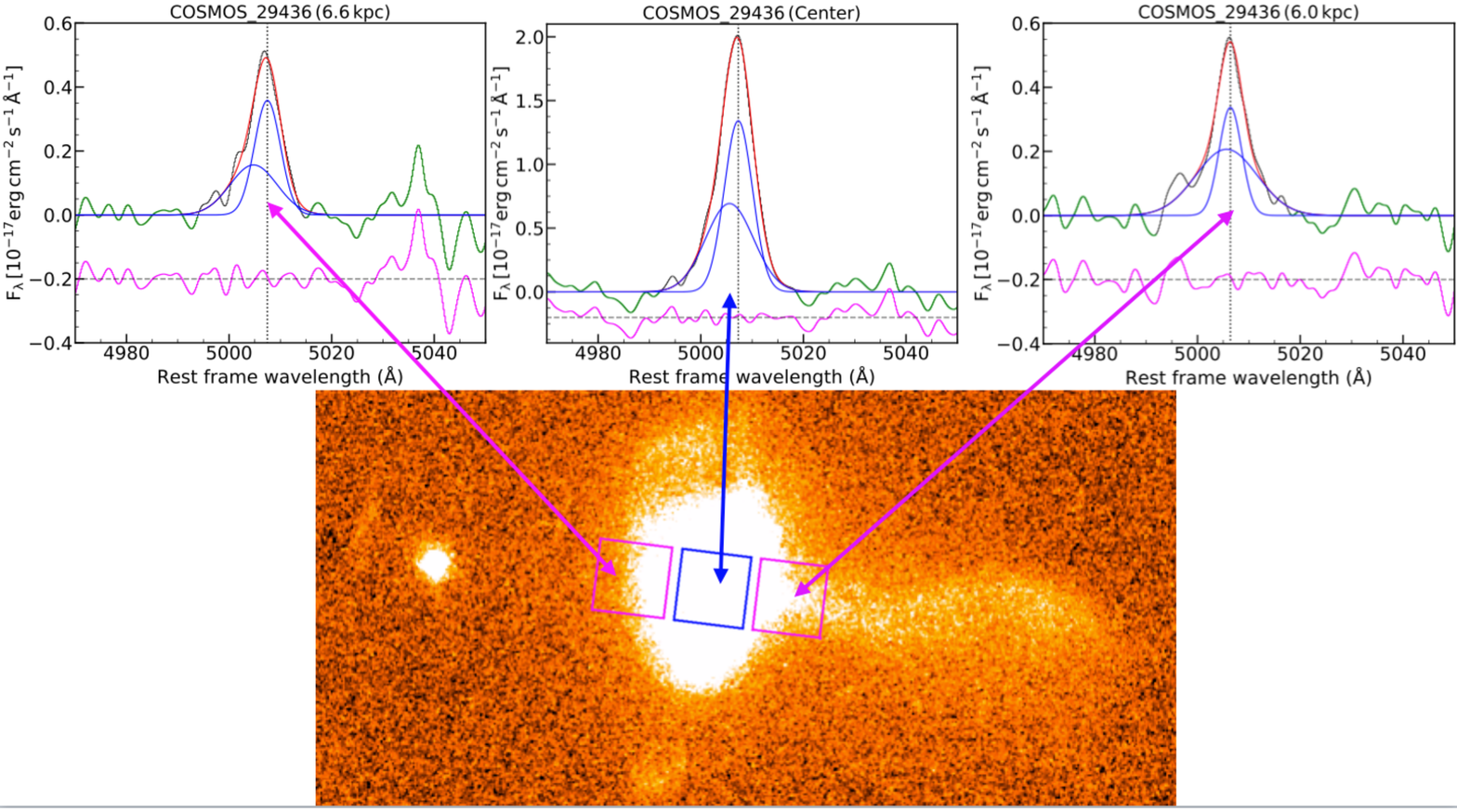}\\
\includegraphics[angle=270,width=.49\textwidth]{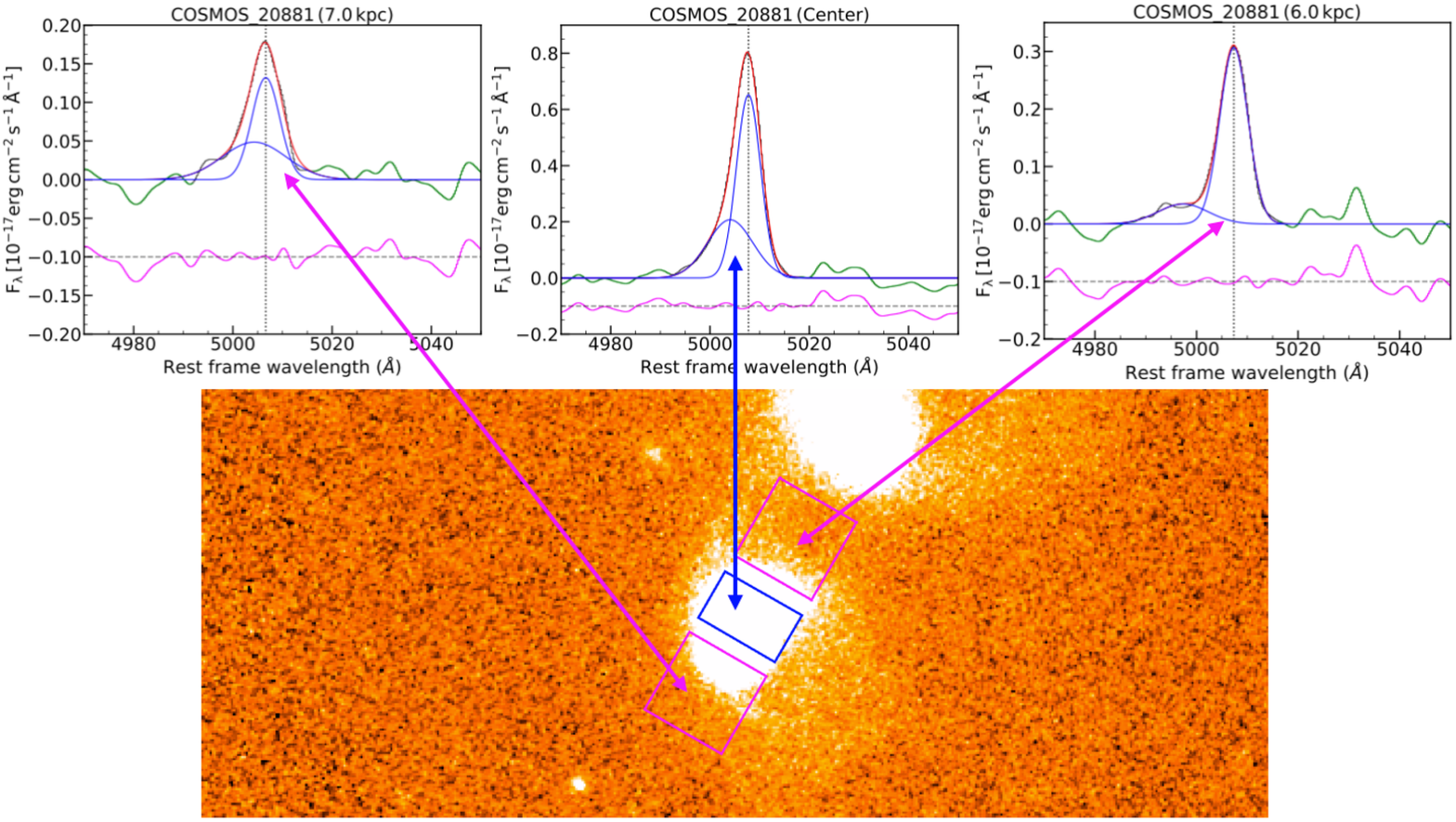}
\includegraphics[angle=270,width=.49\textwidth]{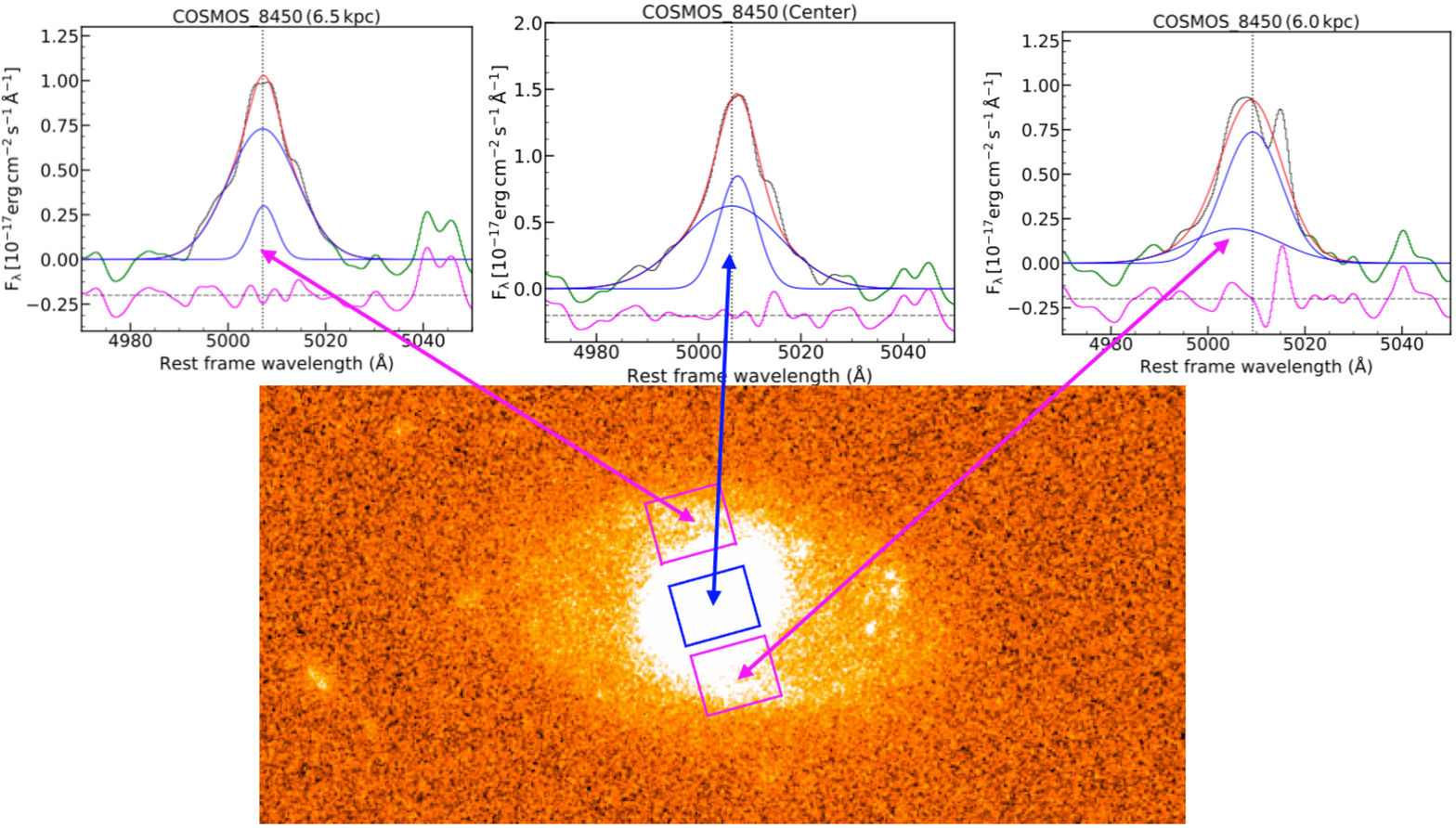}\\
\includegraphics[angle=270,width=.49\textwidth]{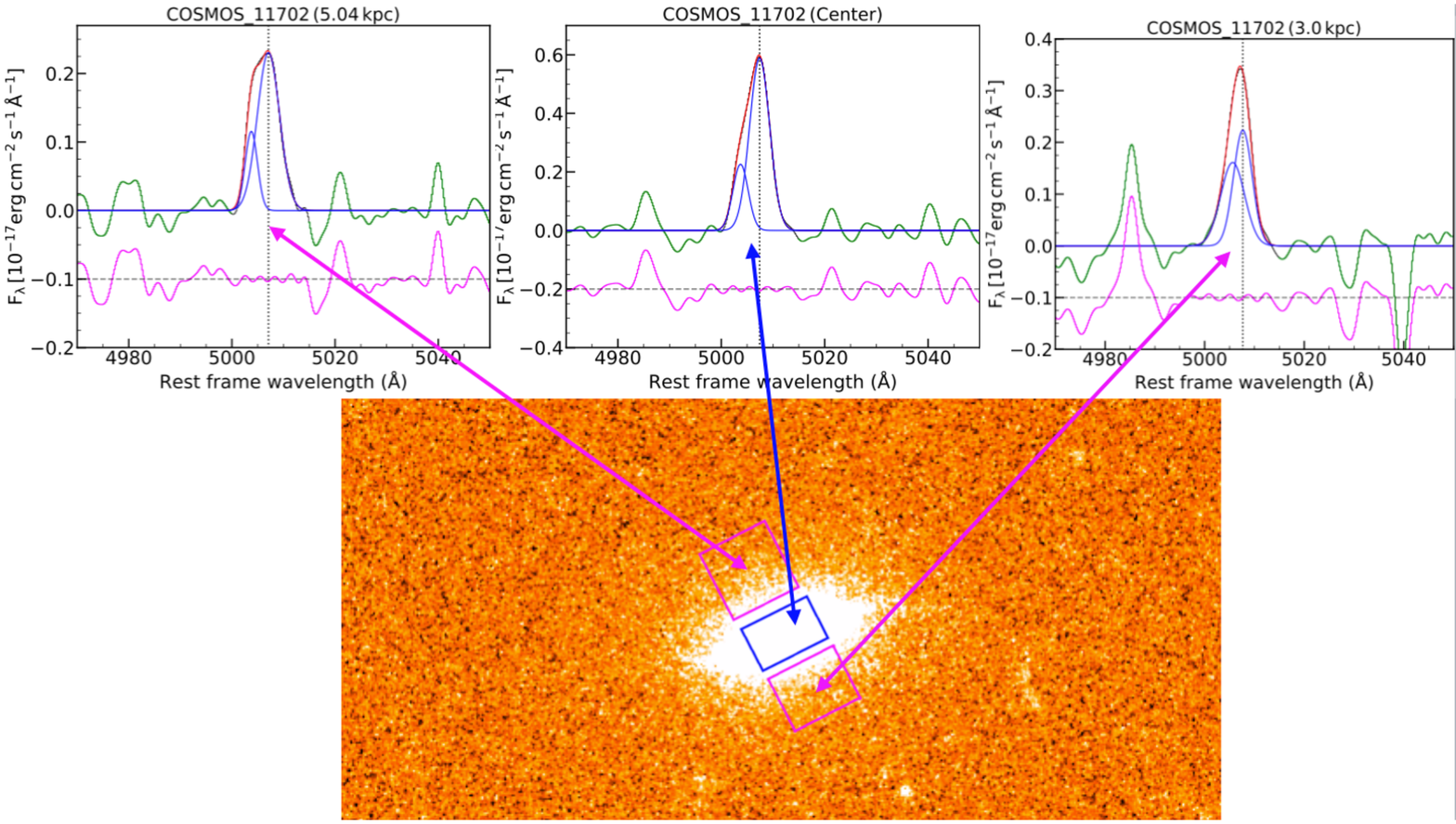}
\caption{For each object, we show the spectral data (in black), the components fitted for emission lines (in blue), the overall total best-fitting model (in red), and the residuals (in magenta).\,The dashed vertical lines mark the location of the rest-frame  [OIII]$\rm{\lambda5007\,\AA}$ line.}
\end{figure*}
\end{document}